\documentclass[journal,10pt,twoside,twocolumn,letterpaper]{IEEEtran}
\ifCLASSOPTIONcompsoc
    \usepackage[caption=false, font=normalsize, labelfont=sf, textfont=sf]{subfig}
\else
\usepackage[caption=false, font=footnotesize]{subfig}
\fi
\usepackage{graphicx, epstopdf}
\usepackage{lipsum}
\usepackage{amssymb}
\usepackage{amsmath}
\usepackage{amsthm}
\usepackage{mathrsfs}
\usepackage{cite}
\usepackage{color}
\usepackage[acronym]{glossaries}
\usepackage[T1]{fontenc}
\usepackage[latin9]{inputenc}
\usepackage{array}
\usepackage{textcomp}
\usepackage{multirow}
\usepackage{flushend}

\newtheorem{theorem}{Theorem}

\newtheorem{remark}{Remark}
\newtheorem{proposition}{Proposition}

\makeatletter
\newcommand*\titleheader[1]{\gdef\@titleheader{#1}}
\AtBeginDocument{%
  \let\st@red@title\@title
  \def\@title{%
    \bgroup\normalfont\small\centering\@titleheader\par\egroup
    \vskip0.5em\st@red@title}
}

\title{Secure Transmit Antenna Selection Protocol for MIMO NOMA Networks over Nakagami-\textit{m} Channels}

\author{ Duc-Dung Tran,~%\IEEEmembership{Member,~IEEE,}
        Ha-Vu Tran,~%\IEEEmembership{Fellow,~OSA,}
		Dac-Binh Ha,
		and~Georges Kaddoum %\IEEEmembership{Life~Fellow,~IEEE}% <-this % stops a space
\thanks{
\textcopyright 2019 IEEE. Personal use of this material is permitted. Permission from IEEE must be obtained for all other uses, in any current or future media, including reprinting/republishing this material for advertising or promotional purposes, creating new collective works, for resale or redistribution to servers or lists, or reuse of any copyrighted component of this work in other works  DOI: {<10.1109/JSYST.2019.2900090>}.}
\thanks{
Duc-Dung Tran and Dac-Binh Ha are with the Faculty of Electrical $\&$ Electronics Engineering, Duy Tan University, Danang, Vietnam (e-mail: tranducdung@dtu.edu.vn, hadacbinh@duytan.edu.vn)}% <-this % stops a space
\thanks{Ha-Vu Tran and Georges Kaddoum are with LACIME Laboratory, ETS Engineering School, University of Quebec, Montreal, Canada (e-mail: ha-vu.tran.1@ens.etsmtl.ca, georges.kaddoum@etsmtl.ca).}% <-this % stops a space
}

\begin{document}
\makeatother
\titleheader{This is the authors' version of the paper that has been accepted for publication in IEEE Systems Journal.}
 \maketitle
 
\begin{abstract}
In this paper, we {consider} a multi-input multi-output (MIMO) non-orthogonal multiple access (NOMA) network consisting of one source and two legitimate users (LUs), so-called near and far users according to their distances to the source, and one passive eavesdropper, over Nakagami-\textit{m} fading channel{s}. Specifically, we investigate the {cases where} the signals of the far user might or might not be {successfully decoded} at the eavesdropper and the near user. Thus, we {aim to design} a transmit antenna selection (TAS) secure communication protocol for the network{; where}, two TAS solutions, namely Solutions I and II, {are} proposed. Specifically, solutions I and II focus on maximizing the received signal power between the source and the near user, and between the source and the far user, respectively. Accordingly, exact and asymptotic closed-form expressions {for} the secrecy outage probability {of} the LUs and the overall system {are} derived. Our analytical results corroborated by Monte Carlo simulation indicate that the secrecy performance could be significantly improved by properly selecting the power allocation coefficients and increasing the number of antennas at the source and the LUs. Interestingly, solution II {is shown to} provide a better overall secrecy performance over solution I.
\end{abstract}


\section{Introduction}
Non-orthogonal multiple access (NOMA) has been recently considered as a promising solution for the next generation of wireless communication networks (i.e., 5G) to improve the spectral efficiency and user fairness \cite{Sai2013, Dai2015}. The principle of NOMA relies on utilizing the power domain to serve multiple users at the same time/frequency/code \cite{Dai2015}. Specifically, in a NOMA network with more than two users, a base station (BS) transmits the {superimposed messages} with different power levels to {the} users \cite{Din2014, Dungwcnc}. The power allocation is carried out based on the users' channel conditions. 
{At the receiver side, successive interference cancellation (SIC), a multiuser data detection technique, is performed to decode the messages. In principle, 
first, a user detects the messages of other users having weaker channel conditions 
and then removes them from observation.} {Second, the user then} decodes its own messages by treating the messages of the other {users} (i.e., the users with stronger channel conditions) as noise. {It becomes evident that NOMA can significantly reduce the multiple access interference at an affordable complexity cost \cite{Liu20171}.}
{Further, in comparison with conventional multiple access schemes such as orthogonal frequency-division multiple access (OFDMA) \cite{Sai2013, Gho2018}, NOMA can provide better user fairness because the users with weak channel conditions are allocated more power.}

With the development of wireless networks, information security is a critical challenge due to the broadcast nature of wireless transmissions \cite{Wyn1975, Csi1978}. Recently, physical layer security (PLS) has been emerging as a prominent candidate for improving the secrecy performance of wireless systems \cite{Bas2011, Tra2015, Tra20152}. The key idea of PLS relies on exploiting the characteristics of wireless channels to guarantee secure communications \cite{Csi1978}. This approach is different from traditional security solutions, such as cryptographic protocols in the upper layers \cite{Sha1949}. In PLS, it is well-known that the perfect secrecy is achieved when the quality of the legitimate channel is higher than that of illegitimate channel \cite{Wyn1975, Csi1978, Bas2011, Tra2015, Tra20152}. To obtain further significant improvements, many PLS-based transmission methods have been proposed, based on the applications of multi-input multi-output (MIMO) \cite{Yan2013}, artificial noises \cite{Goe2008}, jamming \cite{Vo2017}, or beamforming techniques \cite{Muk2011}.

Nowadays, {PLS of NOMA systems} is becoming an interesting topic \cite{Zha2016, Qin2016, Liu2017, Lei2017, Lv2018}. 
In \cite{Zha2016}, the secrecy sum rate of a single-input single-output (SISO) NOMA system has been studied and the closed-form optimal solution {for the} power allocation has been derived, based on the users' quality of service (QoS) requirements.
Taking the PLS of large-scale NOMA networks into account, \cite{Qin2016} has employed stochastic geometry methods to locate users and eavesdroppers in the system. Furthermore, a protected zone around the source has been introduced to enhance the information security. To characterize the secrecy performance of the proposed system, new exact and asymptotic expressions of the secrecy outage probability {(SOP)} have been derived. In \cite{Liu2017}, the application of multiple antennas, artificial noises (AN) and a protected zone to NOMA in large-scale networks have been studied for the purpose of improving the secrecy performance. The obtained results regarding the secrecy outage probability indicated that a significant secrecy performance gain could be achieved by generating the AN at the BS and invoking the protected zone.
In addition, the PLS in a two-user multi-input single-output (MISO) NOMA system was examined in \cite{Lei2017} where NOMA is performed based on the users' QoS requirements, instead of their channel conditions. In this context, the authors considered a scenario that the eavesdropper detects two-user data using SIC. Specifically, it first treats the message of the user with low QoS as noise to decode the message of the user with high QoS, and then subtracts this component from its observation before decoding the message of the {other} user. {Moreover, the use of a transmit antenna selection (TAS) scheme for improving secrecy performance was considered. Besides, with a multi-antenna source, using TAS can help to improve the power efficiency which is a critically important issue to reduce the burden of electrical grids in practice \cite{Mon2017}}. In \cite{Lv2018}, a new secrecy beamforming (SBF) scheme for MISO NOMA systems was studied. Specifically, the proposed SBF scheme efficiently exploited AN to protect the confidential information of two NOMA assisted legitimate users (LUs). {It is clear that the TAS technique has been demonstrated as a low-complexity and power-efficient solution for secure NOMA networks \cite{Lei2017, Chen2017}.} 

Nevertheless, previous works on PLS in NOMA systems \cite{Zha2016, Qin2016, Liu2017, Lei2017, Lv2018} have not covered the following issues.
First, they have considered the scenario of SISO or MISO systems with Rayleigh fading channels. Thus, the applications of MIMO which can significantly improve the secrecy performance and the consideration of Nakagami-$m$ fading channel, {a more general scenario}, have not been addressed yet. 
Second, the works \cite{Zha2016, Qin2016, Liu2017} have been carried out under the ideal assumptions as follows: {\it (i)} the strong user always successfully decodes the message of the weak user, and {\it (ii)} the eavesdroppers have powerful detection abilities to distinguish a multi-user data stream without the interference, generated by superposed transmit signals (i.e., worst-case eavesdropping scenario (WcES)). Unlike these studies, the authors in \cite{Lei2017} have investigated the scenario that the eavesdropper is affected by the interference generated by superposed signals when carrying out a multi-user data detection as discussed above. However, they have assumed that the message of the user with high QoS is already decoded successfully at the eavesdropper when analyzing the secrecy performance of the user with low QoS. Meanwhile, \cite{Lv2018} has considered both transmission outage\footnote{In a two-user NOMA system, the transmission outage occurs at the strong user when it does not successfully decode the message of the weak user or its own message.} and secrecy outage for analyzing the secrecy performance {but only for the WcES scenario}. In fact, {although the WcES is an effective approach to characterize the secrecy performance, this can overestimate the eavesdropper's multi-user decodability.}

Motivated by the aforementioned issues, in this paper, we {propose} a new TAS-based\footnote{{In our TAS method, only one antenna from multiple transmit antennas is selected for transmissions. Choosing multiple antennas, which is more challenging, will be an interesting problem to consider in future work.}} secrecy communication protocol for two-user\footnote{In order to focus on designing a new secrecy communication protocol, a two-user NOMA network is studied in this paper. In particular, our obtained results can be easily used for further calculations in downlink NOMA systems with multiple users (more than two users) by applying the hybrid multiple access techniques (i.e., the combination between NOMA and conventional OMA schemes) as studied in \cite{Din2016, Din20163}.} MIMO NOMA networks over Nakagami-{\it m} fading channels. 
It is assumed that maximal ratio combining (MRC) is employed at the two LUs and the eavesdropper to maximize the signal quality. 
On this basis, two solutions of TAS to design a secure communication protocol, namely Solutions I and II, {are} proposed to maximize the received signal power at the near and far users, respectively. 
To analyze the secrecy performance of the two solutions, we derived the {exact and approximate} closed-form expressions of the SOP for the two LUs and the overall system. Also, we provided the asymptotic expressions for the SOP and investigate the secrecy diversity.
Moreover, we compare the secrecy performance in various scenarios adopting the two solutions, such as SISO versus MIMO, and compared our solutions with the previous works \cite{Liu2017, Lei2017, Lv2018} to evaluate the benefits of our proposed protocol. Accordingly, to validate the analytical results, Monte Carlo simulation was employed.

{Therefore, the contributions of this paper can be summarized as follows:}

\begin{itemize}
	\item {We analyze} the secrecy performance of the NOMA network in which the impact of MIMO transmission, Nakagami-$m$ fading, and the eavesdropper's multi-user decodability on the system performance are addressed.
	\item {We propose} two new TAS protocols to improve the security of the considered MIMO NOMA network.
	\item {We derive exact closed-form expressions for the secrecy outage probability (SOP) of the near user.}
	\item {We derive approximate closed-form expressions for the SOP of the far user as well as for the overall network.}
	\item {We provide} asymptotic expressions for the SOPs and a secrecy diversity order analysis. {Thus, we evaluate the effects of various system parameters, such as average transmit signal-to-noise ratio (SNR), the number of antennas, power allocation coefficients, and fading parameters, on the secrecy performance.}
\end{itemize}

The remainder of the paper is organized as follows: The system model is presented in Section II. The transmit antenna selection scheme is described in Section III. The secrecy performance analysis of the considered system is shown in Section IV. The numerical results and discussions are depicted in Section V. Finally, Section VI shows our conclusion.

\section{System Model \label{sys_model}}
\begin{figure}[!t]
\centering
\includegraphics[scale = 0.35]{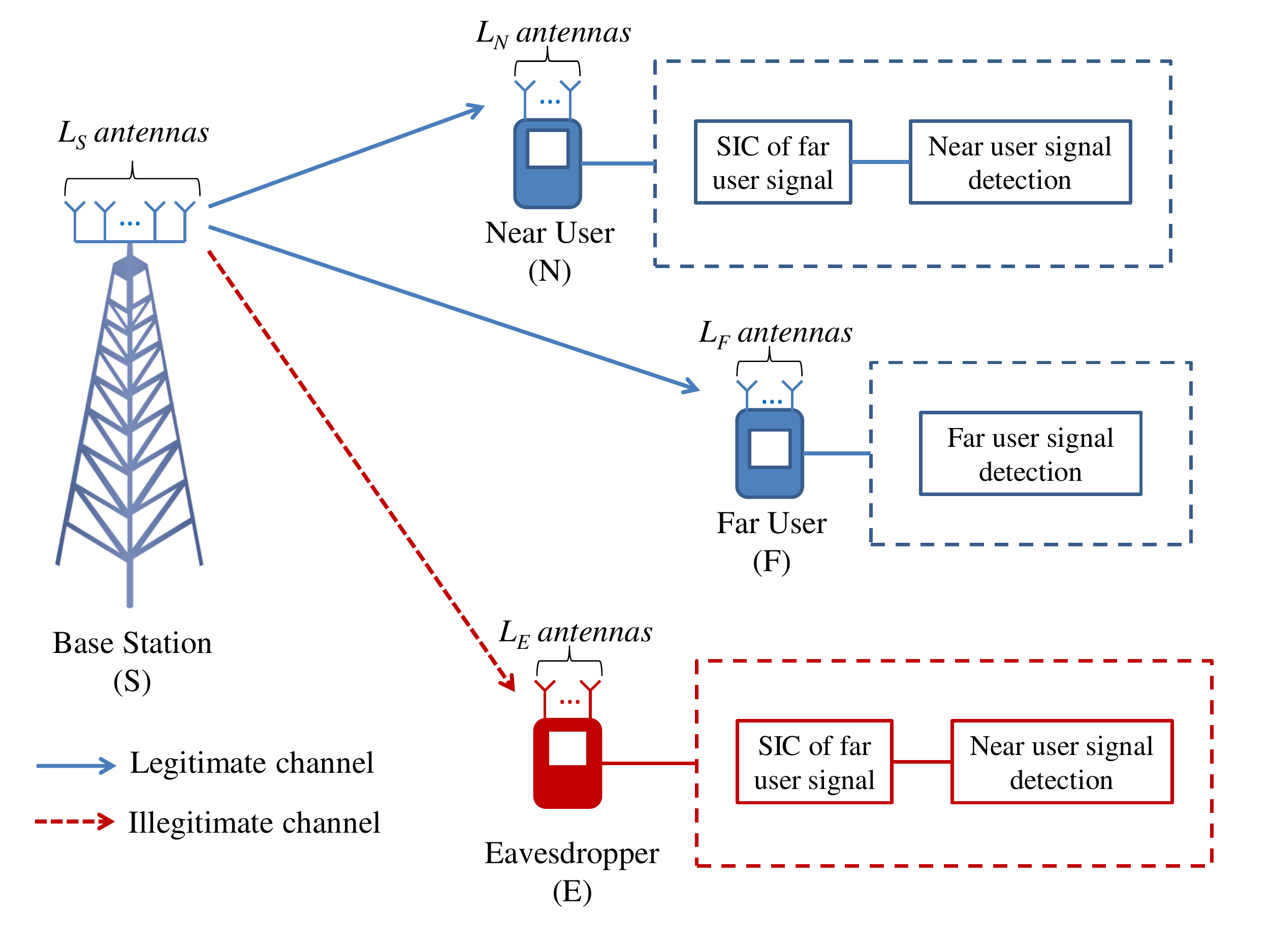}
\caption{System model for secure transmission in MIMO NOMA networks.} 
\label{fig_model}
\end{figure}

Consider a downlink MIMO NOMA network consisting of a source (the base station) denoted by $S$, a near user denoted by $N$, a far (cell edge) user denoted by $F$, and a passive eavesdropper denoted by $E$, as depicted in Fig. \ref{fig_model}. In this system, the source $S$, the near user $N$, the far user $F$, and the eavesdropper $E$ are equipped with $L_S$, $L_N$, $L_F$, and $L_E$ antennas, respectively. 

Let $h_{U_jS_i}$ ($1 \le i \le L_{S}$, $1 \le j \le L_{U}$, $U \in \left\{N, F\right\}$) denote the channel fading coefficient from antenna $i$ at $S$ to antenna $j$ at $U$. 
Similarly, $h_{E_j S_i}$ ($1 \le j \le L_{E}$) denotes the channel fading coefficient from antenna $i$ at $S$ to antenna $j$ at $E$.
In our work, the legitimate and eavesdropping channels are modeled as mutually independent and identically distributed (i.i.d) Nakagami-$m$ fading {channels} with parameters $m_U$ and $m_E$, and squared means $\Omega_U = \mathbb{E}\left[ {\left| h_{{U_j}{S_i}} \right|}^2 \right]$ and $\Omega_E = \mathbb{E}\left[ {\left| h_{{E_j}{S_i}} \right|}^2 \right]$, respectively. In addition, the distance and path loss exponent of {the} $S \to U$ and $S \to E$ channels are denoted by $d_{US}$ and $\theta_{US}$, and $d_{ES}$ and $\theta_{ES}$, respectively.

{Employing NOMA}, $S$ simultaneously communicates with two LUs $N$ and $F$. Further, $S$ selects an antenna among $L_S$ antennas to broadcast information to $N$ and $F$ by applying a TAS technique. The antenna selection schemes will be clarified in the next section. At the receiver side, maximal ratio combining (MRC) is used at $N$, $F$, and $E$.

{Suppose that antenna $i$ at $S$ is selected for transmission, the channel power gain of the $S \to U$ link with MRC can be expressed as}
\begin{equation}
\left\| {{{\bf{h}}_{U{S_i}}}} \right\|^2 = \sum\limits_{j = 1}^{{L_U}} {{\left| {{h_{{U_j}{S_i}}}} \right|}^2},
\label{h2u_mrc}
\end{equation}
where ${\bf{h}}_{US_i} \in \mathbb{C}^{L_U \times 1}$ denotes the channel coefficient vector of {the} $S \to U$ link.

%\subsection{Signal Model}
Given the above discussion, the overall communication process of the system can be mathematically depicted as follows. Following the principle of NOMA, $S$ transmits the superposed message $x = \sqrt {{\alpha _N}{P_S}}x_N + \sqrt {{\alpha _F}{P_S}}x_F$ to $N$ and $F$, where $x_N$ and $x_F$ denote the intended messages for $N$ and $F$, respectively. Also $\alpha_N$ and $\alpha_F$ denote {the} power allocation coefficients for $N$ and $F$, respectively. Thus, the received signal at $U$ ($U \in \left\{N, F\right\}$) is given by
\begin{equation}
y_{US_i} = \left(\sqrt {{\alpha _N}{P_S}} {x_N} + \sqrt {{\alpha _F}{P_S}} {x_F} \right){\bf{w}}_{US_i}{\bf{h}}_{US_i} + n_{US_i},
\label{yu}
\end{equation}
where ${\bf{w}}_{US_i} = \frac{{\bf{h}}_{US_i}^\dag}{\left\| {{\bf{h}}_{US_i}} \right\|}$ represents the signal processing operation at $U$ with MRC, $n_{US_i} \sim {\cal{CN}}(0, N_0)$ stands for an additive white Gaussian noise (AWGN) at user $U$. According to the principle of NOMA, we assume that $\left\| {\bf{h}}_{NS_i} \right\|^2 > \left\| {\bf{h}}_{FS_i} \right\|^2$, $\alpha _F > \alpha _N > 0$ and $\alpha _F + \alpha _N = 1$. Therefore, the instantaneous signal-to-interference-and-noise ratio (SINR) at $F$ to detect $x_F$ is 
\begin{equation}
\gamma _{FS_i}^{x_F} = \frac{\alpha_F \gamma_0 {\left\| {\bf{h}}_{FS_i} \right\|}^2}{\alpha_N \gamma_0 {{\left\| {\bf{h}}_{FS_i} \right\|}^2} + 1},
\label{gfsi}
\end{equation}
where $\gamma_0 = \frac{P_S}{N_0}$ denotes the average transmit signal-to-noise ratio (SNR) associated with the LUs.

At $N$, a SIC receiver is used to decode $x_F$ which is then removed from the observation before detecting $x_N$. Thus, the instantaneous SINR at $N$ to detect $x_F$ can be given by
\begin{equation}
\gamma _{NS_i}^{x_F} = \frac{\alpha _F \gamma _0 {{\left\| {\bf{h}}_{NS_i} \right\|}^2}}{{\alpha _N}{\gamma _0}{{\left\| {\bf{h}}_{NS_i} \right\|}^2} + 1},
\label{gnsi_xf}
\end{equation}
and the instantaneous SNR at $N$ to detect $x_N$ is written as
\begin{equation}
\gamma _{NS_i}^{x_N} = \alpha_N \gamma _0 {\left\| {\bf{h}}_{NS_i} \right\|^2}.
\label{gnsi_sn}
\end{equation}

At $E$, the signal received from $S$ can be expressed as
\begin{equation}
y_{ES_i} = \left( \sqrt {\alpha _N P_S} x_N + \sqrt {\alpha_F P_S} x_F \right){{\bf{w}}_{ES_i}}{{\bf{h}}_{ES_i}} + n_{ES_i},
\label{ye}
\end{equation}
where ${\bf{w}}_{ES_i} = \frac{{\bf{h}}_{ES_i}^\dag}{\left\| {{\bf{h}}_{ES_i}} \right\|}$ represents the signal processing operation at $E$ with MRC, $n_{ES_i} \sim {\cal{CN}}(0, N_E)$ stands for the AWGN at $E$. We assume that SIC receiver is applied at $E$ to decode $x_F$, then subtract this element from the received signal to detect $x_N$. Thus, the instantaneous SINR at $E$ to detect $x_F$ can be expressed as
\begin{equation}
\gamma _{ES_i}^{x_F} = \frac{\alpha _F \gamma _E {{\left\| {{\bf{h}}_{ES_i}} \right\|}^2}}{\alpha _N \gamma _E {{\left\| {{\bf{h}}_{ES_i}} \right\|}^2} + 1},
\label{gesi_xf}
\end{equation}
and the instantaneous SNR at $E$ to detect $x_N$ is written as
\begin{equation}
\gamma _{ES_i}^{x_N} = \alpha _N \gamma _E {\left\| {{\bf{h}}_{ES_i}} \right\|^2},
\label{gesi_xn}
\end{equation}
where $\gamma_E = \frac{P_S}{N_E}$ is the average transmit SNR associated with $E$.

\section{Proposed TAS Protocol and Preliminaries} \label{tas}
In this section, we propose two solutions, namely Solutions I and II, for designing a TAS-based secure protocol for the {considered} MIMO NOMA network. {To do this, it is assumed that the channel state information (CSI) of the legitimate channels is available at $S$, whereas the eavesdropper's CSI is unknown at $S$ \cite{Zhu2016, Lei20171}. {Furthermore, we assume that perfect CSI\footnote{{In fact, the perfect CSI may be difficult to obtain due to the effect of delay and time-varying of the wireless link \cite{Hua2014}. This can be resolved by investigating the imperfect CSI which is an interesting issue to analyze in future work.}} is considered in this paper, as used in \cite{Yan2013, Zha2016, Qin2016, Liu2017, Lei2017, Lv2018}.}}

\subsection{Two Proposed TAS Solutions}
For convenience, let $X_i = {{\left\| {\bf{h}}_{FS_i} \right\|}^2}$, $Y_i = {{\left\| {\bf{h}}_{NS_i} \right\|}^2}$, and $Z_i = {{\left\| {\bf{h}}_{ES_i} \right\|}^2}$.

\subsubsection{{Solution I and Formulations of Legitimate Channels}}
In solution I, an antenna at $S$ is selected for transmission with the aim of finding the best channel condition of {the} $S \to N$ link.

Given this context, the selected antenna, denoted by $\hat i$, can be mathematically represented as follows:
\begin{equation}
\hat i = \arg \mathop {\max }\limits_{1 \le i \le L_S} \left\{ Y_i \right\}.
\label{itas_I}
\end{equation}

With this setting, the CDF of $Y_{\hat i}$ has the following form \cite{Yan2013}
\begin{equation}
F_{Y_{\hat i},I}\left( x \right) = {\left( {1 - \sum\limits_{k = 0}^{a_N - 1} {\frac{{m_N^k}}{{k!\lambda _{NS}^k}}{x^k}{e^{ - \frac{{m_N}x}{\lambda _{NS}}}}} } \right)^{{L_S}}},
\label{cdf_hnsI_def}
\end{equation}
where ${a_N} = {m_N}{L_N}$ and $\lambda _{NS} = \frac{\Omega_N}{d_{NS}^{\theta _{NS}}}$.

With the aid of the binomial expansion given in \cite[Eq. 1.111]{Grad2007} and using the multinomial theorem, (\ref{cdf_hnsI_def}) can be rewritten as
\begin{equation}
F_{Y_{\hat i},I}\left( x \right) = 1 + \sum\limits_{p = 1}^{{L_S}} {\sum\limits_{{\Delta _N} = p} {{\Phi _N}{x^{{\varphi _N}}}{e^{ - \frac{{p{m_N}x}}{{{\lambda _{NS}}}}}}} },
\label{cdf_hnsI_final}
\end{equation}
where ${\Delta _N} = \sum\limits_{q = 0}^{{a_N - 1}} {{\delta _{N,q}}}$, $\varphi_N = \sum\limits_{q = 0}^{a_N - 1} {q{\delta _{N,q}}}$, and ${\Phi _N} = \left( {\begin{array}{*{20}{c}}
{{L_S}}\\
p
\end{array}} \right){\left( { - 1} \right)^p}\left( {\begin{array}{*{20}{c}}
p\\
{{\delta _{N,0}},...,{\delta _{N,{a_N - 1}}}}
\end{array}} \right)\left[ {\prod\limits_{q = 0}^{{a_N - 1}} {{{\left( {{\frac{{m_N^q}}{{q!\lambda _{NS}^q}}}} \right)}^{{\delta _{N,q}}}}} } \right]$.

{Moreover}, the CDF of $X_{\hat i}$ is given by \cite{Yan2013}
\begin{equation}
F_{X_{\hat i},I}\left( x \right) = 1 - \sum\limits_{k = 0}^{{a_F - 1}} {\frac{m_F^k}{{k!\lambda _{FS}^k}}{x^k}{e^{ - \frac{{{m_F}x}}{{{\lambda_{FS}}}}}}},
\label{cdf_hfsI_final}
\end{equation}
where $a_F = m_F L_F$ and $\lambda _{FS} = \frac{\Omega_F}{d_{FS}^{\theta_{FS}}}$.

\subsubsection{{Solution II and Formulations of Legitimate Channels}}
{In contrast to} Solution I, in Solution II, an antenna at $S$ is chosen to broadcast information with the purpose of providing the best channel gain of {the} $S \to F$ link. Accordingly, the chosen antenna at $S$ in this case can be expressed as
\begin{equation}
\hat i = \arg \mathop {\max }\limits_{1 \le i \le L_S} \left\{ {X_i} \right\}.
\label{itas_II}
\end{equation}

Thus, the CDF of $X_{\hat i}$ can be given as
\begin{equation}
F_{X_{\hat i},II}\left( x \right) = 1 + \sum\limits_{p = 1}^{{L_S}} {\sum\limits_{{\Delta _F} = p} {{\Phi _F}{x^{{\varphi _F}}}{e^{ - \frac{{p{m_F}x}}{{{\lambda _{FS}}}}}}} },
\label{cdf_hfsII_final}
\end{equation}
and the CDF of $Y_{\hat i}$ is 
\begin{equation}
F_{Y_{\hat i},II}\left( x \right) = 1 - \sum\limits_{k = 0}^{{a_N - 1}} {\frac{m_N^k}{{k!\lambda_{NS}^k}}{x^k}{e^{ - \frac{{{m_N}x}}{{{\lambda_{NS}}}}}}},
\label{cdf_hnsII_final}
\end{equation}
where ${\Delta _F} = \sum\limits_{q = 0}^{{a_F - 1}} {{\delta _{F,q}}}$, $\varphi_F = \sum\limits_{q = 0}^{a_F - 1} {q{\delta _{F,q}}}$, and ${\Phi _F} = \left( {\begin{array}{*{20}{c}}
{{L_S}}\\
p
\end{array}} \right){\left( { - 1} \right)^p}\left( {\begin{array}{*{20}{c}}
p\\
{{\delta _{F,0}},...,{\delta _{F,{a_F - 1}}}}
\end{array}} \right)\left[ {\prod\limits_{q = 0}^{{a_F - 1}} {{{\left( {{\frac{{m_F^q}}{{q!\lambda _{FS}^q}}}} \right)}^{{\delta _{F,q}}}}} } \right]$.

\subsubsection{Formulations of Eavesdropping Channels}
For $E$, the PDF and the CDF of $Z_{\hat i}$ in both Solutions I and II are given by \cite{Yan2013}
\begin{equation}
{f_{{Z_{\hat i}}}}\left( x \right) = \frac{m_E^{a_E}{x^{{a_E - 1}}}}{{\Gamma \left( {a_E} \right)\lambda _{ES}^{a_E}}}{e^{ - \frac{{{m_E}x}}{{{\lambda _{ES}}}}}},
\label{pdf_hes_final}
\end{equation}
\begin{equation}
F_{Z_{\hat i}}\left( x \right) = 1 - \sum\limits_{k = 0}^{{a_E - 1}} {\frac{m_E^k}{{k!\lambda _{ES}^k}}{x^k}{e^{ - \frac{{{m_E}x}}{{{\lambda _{ES}}}}}}},
\label{cdf_hes_final}
\end{equation}
where $a_E = m_E L_E$ and $\lambda_{ES} = \frac{\Omega_E}{d_{ES}^{\theta_{ES}}}$.

\subsection{Preliminary Analysis of SINRs with Solutions I and II}
From (\ref{cdf_hfsI_final}), (\ref{cdf_hfsII_final}), and (\ref{cdf_hes_final}), the closed-form expressions of the CDF of $\gamma_{FS_{\hat i}}^{x_F}$ in Solutions I and II, and the PDF of $\gamma_{ES_{\hat i}}^{x_F}$ are {derived in Proposition 1, Proposition 2, and Proposition 3}, respectively, as follows.

\begin{proposition}
Under Nakagami-$m$ fading, the CDF of $\gamma_{FS_{\hat i}}^{x_F}$ in Solution I has the following form
\begin{equation}
\begin{split}
&F_{{\gamma_{FS_{\hat i}}^{x_F}},I}\left( x \right)\\
& = \left\{ {\begin{array}{*{20}{c}}
{1,}&{x \ge \beta }\\
{1 - \sum\limits_{k = 0}^{a_F - 1} {\frac{1}{k!\lambda_{FS}^k}{{\left( {\frac{{m_F}{A_x}}{\gamma_0}} \right)}^k}{e^{ - \frac{{m_F}{A_x}}{{\gamma_0}{\lambda _{FS}}}}}} ,}&{x < \beta }
\end{array}} \right.,
\end{split}
\label{cdf_gfsxfI_final}
\end{equation}
where $A_x = \frac{x}{{\alpha_F} - {\alpha_N}x}$ and $\beta = \frac{\alpha_F}{\alpha_N}$.
\end{proposition}

\begin{IEEEproof}
With the aid of (\ref{gfsi}), $F_{{\gamma_{FS_{\hat i}}^{x_F}},I}\left( x \right)$ is given by
\begin{equation}
\begin{split}
{F_{\gamma _{F{S_{\hat i}}}^{{x_F}},I}}\left( x \right) &= \Pr \left( {\frac{{{\alpha _F}{\gamma _0}{X_{\hat i}}}}{{{\alpha _N}{\gamma _0}{X_{\hat i}} + 1}} < x} \right) \\
%& = \left\{ {\begin{array}{*{20}{c}}
%{1,}&{x \ge \beta }\\
%{\Pr \left( {{X_{\hat i}} < \frac{{{A_x}}}{{{\gamma _0}}}} \right),}&{x < \beta }
%\end{array}} \right. \\
& = \left\{ {\begin{array}{*{20}{c}}
{1,}&{x \ge \beta }\\
{{F_{{X_{\hat i}},I}}\left( {\frac{{{A_x}}}{{{\gamma _0}}}} \right),}&{x < \beta }
\end{array}} \right..
\end{split}
\label{cdf_gfsxfI_cal}
\end{equation}
By substituting (\ref{cdf_hfsI_final}) into (\ref{cdf_gfsxfI_cal}), (\ref{cdf_gfsxfI_final}) is obtained and the proof is completed.
\end{IEEEproof}

\begin{proposition}
Under Nakagami-$m$ fading, the CDF of $\gamma_{FS_{\hat i}}^{x_F}$ in Solution II is expressed as
\begin{equation}
\begin{split}
&{F_{\gamma _{F{S_{\hat i}}}^{{x_F}},II}}\left( x \right)\\
& = \left\{ {\begin{array}{*{20}{c}}
{1,}&{x \ge \beta }\\
{1 + \sum\limits_{p = 1}^{{L_S}} {\sum\limits_{{\Delta _F} = p} {{\Phi _F}{{\left( {\frac{{{A_x}}}{{{\gamma _0}}}} \right)}^{{\varphi _F}}}{e^{ - \frac{{p{m_F}{A_x}}}{{{\gamma _0}{\lambda _{FS}}}}}}} } ,}&{x < \beta }
\end{array}} \right..
\end{split}
\label{cdf_gfsxfII_final}
\end{equation}
\end{proposition}

\begin{IEEEproof}
Similar to the proof of Lemma 1, based on (\ref{gfsi}), $F_{{\gamma_{FS_{\hat i}}^{x_F}},II}\left( x \right)$ can be represented as
\begin{equation}
{F_{\gamma _{F{S_{\hat i}}}^{{x_F}},II}}\left( x \right) = \left\{ {\begin{array}{*{20}{c}}
{1,}&{x \ge \beta }\\
{{F_{{X_{\hat i}},II}}\left( {\frac{{{A_x}}}{{{\gamma _0}}}} \right),}&{x < \beta }
\end{array}} \right..
\label{cdf_gfsxfII_cal}
\end{equation}
To this end, by substituting (\ref{cdf_hfsII_final}) into (\ref{cdf_gfsxfII_cal}) to obtain (\ref{cdf_gfsxfII_final}), the proof is completed.
\end{IEEEproof}

\begin{proposition}
Under Nakagami-$m$ fading, the PDF of $\gamma_{ES_{\hat i}}^{x_F}$ is given by
\begin{equation}
\begin{split}
& f_{\gamma _{ES_{\hat i}}^{x_F}}\left( x \right) \\
& = \left\{ {\begin{array}{*{20}{c}}
{0,}&{x \ge \beta }\\
{\sum\limits_{k = 0}^{{a_E - 1}} {\frac{m_E^k{\alpha_F}{A_x^{k - 1}}{e^{ - \frac{{m_E}{A_x}}{{\gamma _E}{\lambda_{ES}}}}}}{{k!\lambda_{ES}^k\gamma_E^k{{\left( {{\alpha_F} - {\alpha_N}x} \right)}^2}}}\left( {\frac{{m_E}{A_x}}{{\lambda_{ES}}{\gamma_E}} - k} \right)} ,}&{x < \beta }
\end{array}} \right..
\end{split}
\label{pdf_gesxf_final}
\end{equation}
\end{proposition}

\begin{IEEEproof}
First, we derive the CDF of $\gamma_{ES_{\hat i}}^{x_F}$ by using (\ref{gesi_xf}) as follows:
\begin{equation}
{F_{\gamma _{E{S_{\hat i}}}^{{x_F}}}}\left( x \right) = \left\{ {\begin{array}{*{20}{c}}
{1,}&{x \ge \beta }\\
{{F_{{Z_{\hat i}}}}\left( {\frac{{{A_x}}}{{{\gamma_E}}}} \right),}&{x < \beta }
\end{array}} \right..
\label{cdf_gesxf_cal}
\end{equation}
By substituting (\ref{cdf_hes_final}) into (\ref{cdf_gesxf_cal}), ${F_{\gamma _{E{S_{\hat i}}}^{{x_F}}}}\left( x \right)$ is expressed as
\begin{equation}
{F_{\gamma _{E{S_{\hat i}}}^{{x_F}}}}\left( x \right) = \left\{ {\begin{array}{*{20}{c}}
{1,}&{x \ge \beta }\\
{1 - \sum\limits_{k = 0}^{a_E - 1} {\frac{1}{k!\lambda_{ES}^k}{{\left( {\frac{{m_E}{A_x}}{\gamma_E}} \right)}^k}{e^{ - \frac{{m_E}{A_x}}{{\gamma_E}{\lambda _{ES}}}}}} ,}&{x < \beta }
\end{array}} \right..
\label{cdf_gesxf_final}
\end{equation}

The PDF of $\gamma _{E{S_{\hat i}}}^{{x_F}}$ is defined as
\begin{equation}
f_{\gamma _{ES_{\hat i}}^{x_F}}\left( x \right) = \frac{d{F_{\gamma_{ES_{\hat i}}^{x_F}}}\left( x \right)}{dx}.
\label{pdf_gesxf_def}
\end{equation}
The final expression of $f_{\gamma _{ES_{\hat i}}^{x_F}}\left( x \right)$ in (\ref{pdf_gesxf_final}) is obtained by deriving the derivative of ${F_{\gamma _{E{S_{\hat i}}}^{{x_F}}}}\left( x \right)$ in (\ref{cdf_gesxf_final}) with respect to $x$. The proof is completed.
\end{IEEEproof}

\section{Secrecy Performance Analysis}\label{sec_spa}
In this section, to validate the two proposed solutions, the secrecy performance regarding the SOP obtained at $N$ and $F$, and the SOP of the overall system is analyzed.

\subsection{Preliminary}
This subsection presents the definitions of secrecy capacity and SOP for secure communication in the considered MIMO NOMA system.

First, let $C_{U,i}$ and $C_{EU,i}$ $\left(U \in \left\{N, F\right\} \right)$ denote the capacities of {the} $S \to U$ legitimate channel and {the} $S \to E$  illegitimate channel to detect the signal $x_U$, respectively. 
Thus, according to \cite{Blo2008}, the secrecy capacity achieved at $U$ can be defined as
\begin{equation}
\begin{split}
C_{S,i} &=\max\{0,\,C_{U,i}-C_{EU,i}\}\\
&=\max\left\{ 0,\,\log_{2} \left(\dfrac{1+\gamma_{US_i}^{x_U}}{1+\gamma_{ES_i}^{x_U}} \right) \right\},
\end{split}
\label{CSi_def}
\end{equation}
where $C_{U,i}=\log_{2}\left(1+\gamma_{US_i}^{x_U}\right)$ and $C_{EU,i}=\log_{2}\left(1+\gamma_{ES_i}^{x_U}\right)$.

Second, the SOP obtained at $N$ is defined as
\begin{equation}
\begin{split}
&SOP_{N,i}\\
&= {\Pr \left\{ {\gamma _{NS_i}^{x_F} < {\gamma_{th}}} \right\}}\\
&+ {\Pr \left\{ {\gamma _{NS_i}^{x_F} \ge {\gamma_{th}},\gamma _{ES_i}^{x_F} < {\gamma_{th}},C_{N,i} < R_{s,N}} \right\}} \\
 &+ {\Pr \left\{ {\gamma _{NS_i}^{x_F} \ge {\gamma_{th}},\gamma _{ES_i}^{x_F} \ge {\gamma_{th}},C_{S,i} < R_{s,N}} \right\}}, \\
 \end{split}
\label{SOPNi}
\end{equation}
where $\gamma_{th} = 2^{R_F} - 1$ denotes the SNR threshold for correctly decoding $x_F$, $R_F$ represents the target data rate of $F$, and $R_{s,N}$ represents the secrecy rate threshold at $N$. 
 
\begin{remark}
 The definition of $SOP_{N,i}$, given in \eqref{SOPNi}, is different from the previous works considering the scenario that $N$ and $E$ {are always successfully decoding} the message of $F$, i.e., $\Pr \left\{ \gamma _{NS_i}^{x_F} \ge {\gamma_{th}} \right\} = 1$ \cite{Zha2016, Qin2016, Liu2017, Lei2017} and $\Pr \left\{ \gamma _{ES_i}^{x_F} \ge {\gamma_{th}} \right\} = 1$ \cite{Zha2016, Qin2016, Liu2017, Lei2017, Lv2018}.
\end{remark}

To analyze the SOP at user $N$, its formulation needs to be further expressed for more insights.
Thus, according to \eqref{SOPNi}, $SOP_{N,i}$ can be expressed as in {Proposition} 4.

\begin{proposition}
{From \eqref{SOPNi}, the SOP at user $N$ can be further represented as}
\begin{equation}
\begin{split}
&SOP_{N,i}\\
&= \underbrace {\Pr \left\{ {\gamma _{NS_i}^{x_F} < {{\gamma_{th}}}} \right\}}_{\Theta_1}\\
& \hspace{0.3cm} + \underbrace {\Pr \left\{ {\gamma _{NS_i}^{x_F} \ge {\gamma_{th}},\gamma _{ES_i}^{x_F} < {\gamma_{th}},C_{N,i} < R_{s,N}} \right\}}_{\Theta_2} \\
 & \hspace{0.3cm} + \underbrace {\Pr \left\{ {\gamma _{NS_i}^{x_F} \ge {\gamma_{th}},\gamma _{ES_i}^{x_F} \ge {\gamma_{th}},C_{S,i} < R_{s,N}} \right\}}_{\Theta _3} \\
 &= \left\{ {\begin{array}{*{20}{c}}
{1,}&{{{\gamma_{th}}} \ge \beta }\\
{\Lambda_1 + \Lambda_2 + \Lambda_3,}&{{\gamma_{th}} < \beta }
\end{array}} \right.,
 \end{split}
\label{SOPNi_def}
\end{equation}
where $\Lambda_1$, $\Lambda_2$, and $\Lambda_3$ are, respectively, given by
\begin{equation}
\Lambda_1 = {F_{Y_ i}}\left( {\frac{A_{\gamma_{th}}}{\gamma_0}} \right), \nonumber
\end{equation}
\begin{equation}
{\Lambda _2} = \left\{ {\begin{array}{*{20}{c}}
{0,}&{{R_{s,N}} < \eta }\\
{\begin{array}{*{20}{c}}
{\left[ {{F_{{Y_i}}}\left( {\frac{{{\gamma _{s,N}}}}{{{\gamma _0}}}} \right) - {F_{{Y_i}}}\left( {\frac{{{A_{{\gamma_{th}}}}}}{{{\gamma _0}}}} \right)} \right]}\\
{ \times {F_{{Z_i}}}\left( {\frac{{{A_{{\gamma_{th}}}}}}{{{\gamma _E}}}} \right) \hspace{2cm}}
\end{array},}&{{R_{s,N}} > \eta }
\end{array}} \right., \nonumber
\end{equation}
and
\begin{small}\begin{equation}
{\Lambda _3} = \int\limits_{\frac{{{A_{{\gamma_{th}}}}}}{{{\gamma _E}}}}^\infty  {\left[ {{F_{{Y_i}}}\left( {\frac{{{2^{{R_{s,N}}}}{\gamma _E}x + {\gamma _{s,N}}}}{{{\gamma _0}}}} \right) - {F_{{Y_i}}}\left( {\frac{{{A_{{\gamma_{th}}}}}}{{{\gamma _0}}}} \right)} \right]{f_{{Z_i}}}\left( x \right)dx}, \nonumber
\end{equation}
\end{small}where $\eta  = {\log _2}\left( {\frac{\alpha_F}{{\alpha_F} - {\alpha_N}{\gamma_{th}}}} \right)$ and ${\gamma _{s,N}} = \frac{{{2^{{R_{s,N}}}} - 1}}{{{\alpha _N}}}$.
\end{proposition}

\begin{IEEEproof}
See Appendix A.
\end{IEEEproof}

Third, the SOP at $F$ has the following form
\begin{equation}
\begin{split}
SOP_{F,i} &= \Pr \left\{ C_{F,i} - C_{EF,i} < R_{s,F} \right\} \\
&= \int\limits_0^\infty  {{F_{\gamma_{FS_i}^{x_F}}}\left( g_{x,F} \right){f_{\gamma _{ES_i}^{x_F}}}\left( x \right)dx},
\end{split}
\label{SOPFi_def}
\end{equation}
where $g_{x,F} = {2^{R_{s,F}}}x + {2^{R_{s,F}}} - 1$ and $R_{s,F}$ represents the secrecy rate threshold at $F$.

Fourth, in the system, we define the overall SOP as the probability that the secrecy outage event occurs at either $N$ or $F$, i.e.,
\begin{equation}
SOP_{O,i} = 1 - \left( 1 - SOP_{F,i} \right)\left( 1 - SOP_{N,i} \right).
\label{SOPOi_def}
\end{equation}

\subsection{Secrecy Performance Analysis of Solution I}
\subsubsection{{Secrecy Outage Probability Analysis}}
In this case, the SOP at $F$ and $N$ are derived through Theorem 1 and Theorem 2 as follows.

\begin{theorem}
Under Nakagami-$m$ fading, the SOP of user $F$ in Solution I is approximately expressed as
\begin{equation}
SO{P_{F,I}} \approx 1 - \sum\limits_{m = 0}^{{a_F}} {\sum\limits_{n = 0}^{{a_E}} {\sum\limits_{i = 0}^N {\Psi _{F,I}^{(1)}{h_{F,I}}\left[ {\frac{{\left( {{v_i} + 1} \right){u_F}}}{2}} \right]} } },
\label{SOPFI_final}
\end{equation}
where $\Psi _{F,I}^{(1)} = \frac{{\pi {u_F}{\alpha _F}m_F^mm_E^n\sqrt {1 - v_i^2} }}{{m!n!2N\lambda _{FS}^m\lambda _{ES}^n\gamma _0^m\gamma _E^n}}$, $v_i = \cos \left[ {\frac{\left( {2i - 1} \right)\pi}{2N}} \right]$, $u_F = \frac{1}{{{\alpha _N}{2^{{R_{s,F}}}}}} - 1$, $h_{F,I}\left( x \right) = \frac{A_{g_{x,F}}^m A_x^{n - 1}}{{\left( {\alpha _F} - {\alpha _N}x \right)}^2}{e^{ - \frac{{m_F} A_{g_{x,F}}}{{\gamma_0}{\lambda_{FS}}} - \frac{{m_E}{A_x}}{{\gamma_E}{\lambda_{ES}}}}}\left( {\frac{{m_E}{A_x}}{{\gamma_E}{\lambda_{ES}}} - n} \right)$, $g_{x,F} = {2^{R_{s,F}}}x + 2^{R_{s,F}} - 1$, and $N$ is a complexity-accuracy trade-off parameter.
\end{theorem}

\begin{IEEEproof}
See Appendix B.
\end{IEEEproof}

\begin{theorem} 
Under Nakagami-$m$ fading, the SOP of user $N$ in Solution I is given by
\begin{equation}
SOP_{N,I} = \left\{ {\begin{array}{*{20}{c}}
{1,}&{{\gamma_{th}} \ge \beta }\\
{{\Lambda _{1,I}} + {\Lambda _{2,I}} + {\Lambda _{3,I}},}&{{\gamma_{th}} < \beta }
\end{array}} \right.,
\label{SOPNI_final}
\end{equation}
where 
\begin{equation}
\Lambda _{1,I} = {F_{{Y_{\hat i}},I}}\left( {\frac{{{A_{{\gamma_{th}}}}}}{{{\gamma _0}}}} \right), \nonumber
\end{equation}
\begin{equation}
{\Lambda _{2,I}} = \left\{ {\begin{array}{*{20}{c}}
{0,}&{{R_{s,N}} < \eta }\\
{\begin{array}{*{20}{c}}
{\left[ {{F_{{Y_{\hat i}},I}}\left( {\frac{{{\gamma _{s,N}}}}{{{\gamma _0}}}} \right) - {F_{{Y_{\hat i}},I}}\left( {\frac{{{A_{{\gamma_{th}}}}}}{{{\gamma _0}}}} \right)} \right]}\\
{ \times {F_{{Z_{\hat i}}}}\left( {\frac{{{A_{{\gamma_{th}}}}}}{{{\gamma _E}}}} \right) \hspace{2.5cm}}
\end{array},}&{{R_{s,N}} > \eta }
\end{array}} \right., \nonumber
\end{equation}
\begin{equation}
{\Lambda _{3,I}} = B_{N,I}^{(1)} + \sum\limits_{p,{\Delta _N},m,n}^{ \sim}  {{\Psi _{N,I}}{{\left[ {\frac{{{A_{{\gamma_{th}}}}B_{N,I}^{(2)}}}{{{\gamma _E}}}} \right]}^n}{e^{ - \frac{{{A_{{\gamma_{th}}}}B_{N,I}^{(2)}}}{{{\gamma _E}}}}}}, \nonumber
\end{equation}
with $F_{{Y_{\hat i}},I}\left( x \right)$ and $F_{{Z_{\hat i}}}\left( x \right)$ are defined in (\ref{cdf_hnsI_final}) and (\ref{cdf_hes_final}), respectively. $B_{N,I}^{(1)} = \left[ {1 - {F_{{Y_{\hat i}},I}}\left( {\frac{{{A_{{\gamma_{th}}}}}}{{{\gamma _0}}}} \right)} \right]\left[ {1 - {F_{{Z_{\hat i}}}}\left( {\frac{{{A_{{\gamma_{th}}}}}}{{{\gamma _E}}}} \right)} \right]$, $B_{N,I}^{(2)} = \frac{{p{m_N}{\gamma _E}{2^{{R_{s,N}}}}}}{{{\gamma _0}{\lambda _{NS}}}} + \frac{{{m_E}}}{{{\lambda _{ES}}}}$, $\sum\limits_{p,{\Delta _N},m,n}^ \sim = \sum\limits_{p = 1}^{{L_S}} {\sum\limits_{{\Delta _N} = p} {\sum\limits_{m = 0}^{{\varphi _N}} {\sum\limits_{n = 0}^{a_E + m - 1}}}}$, and ${\Psi _{N,I}} = \left( {\begin{array}{*{20}{c}}
{{\varphi _N}}\\
m
\end{array}} \right)\frac{{{\Phi _N}\Gamma \left( {{a_E} + m} \right)m_E^{{a_E}}{{\left( {{2^{{R_{s,N}}}}{\gamma _E}} \right)}^m}\gamma _{s,N}^{{\varphi _N} - m}{e^{ - \frac{{p{m_N}{\gamma _{s,N}}}}{{{\gamma _0}{\lambda _{NS}}}}}}}}{{n!\Gamma \left( {{a_E}} \right)\lambda _{ES}^{{a_E}}\gamma _0^{{\varphi _N}}{{\left[ {B_{N,I}^{(2)}} \right]}^{{a_E} + m}}}}$.
\end{theorem}

\begin{IEEEproof}
See Appendix C.
\end{IEEEproof}

From (\ref{SOPOi_def}), the overall SOP of the system in Solution I is expressed as
\begin{equation}
SOP_{O,I} = 1 - \left( 1 - SOP_{F,I} \right)\left( 1 - SOP_{N,I} \right).
\label{SOPOI_def}
\end{equation}

\subsubsection{Asymptotic Secrecy Outage Probability Analysis \label{Asymp1}}

Using the series representation of $e^x$ given by \cite[Eq. 1.211]{Grad2007}
\begin{equation}
e^x = \sum\limits_{k=0}^{\infty} {\frac{x^k}{k!}},
\label{ex_series}
\end{equation}
the asymptotic CDF of $X_{\hat i}$ and $Y_{\hat i}$ when $\gamma_0 \to \infty$ are written as
\begin{equation}
{F_{{X_{\hat i}},I}}\left( x \right) \approx \frac{{{{\left( {{m_F}x/{\lambda _{FS}}} \right)}^{a_F}}}}{{\left( {a_F} \right)!}},
\label{cdf_hfsI_asym}
\end{equation}
and
\begin{equation}
{F_{{Y_{\hat i}},I}}\left( x \right) \approx {\left[ {\frac{{{{\left( {{m_N}x/{\lambda _{NS}}} \right)}^{a_N}}}}{{\left( {a_N} \right)!}}} \right]^{{L_S}}},
\label{cdf_hnsI_asym}
\end{equation}
respectively.

For $F$, according to {Proposition} 1, (\ref{SOPFi_def}), and (\ref{cdf_hfsI_asym}), after some algebraic manipulations similar to the proof of Theorem 1 in Appendix B, its SOP in Solution I is asymptotically approximated as
\begin{equation}
\begin{split}
SOP_{F,I}^{asym} &\approx  \sum\limits_{m = 0}^{{a_E-1}} {\sum\limits_{i = 0}^N {\Psi _{F,I}^{asym}h_{F,I}^{asym}\left[ {\frac{{\left( {{v_i} + 1} \right){u_F}}}{2}} \right]} } \\
& \quad + \sum\limits_{m = 0}^{{a_E-1}} {\Psi _{F,E}},
\end{split}
\label{SOPFI_asym}
\end{equation}
where $\Psi _{F,E} = \frac{{m_E^mA_{{u_F}}^m{e^{ - \frac{{{m_E}{A_{{u_F}}}}}{{{\gamma _E}{\lambda _{ES}}}}}}}}{{m!\gamma _E^m\lambda _{ES}^m}}$, $\Psi _{F,I}^{asym} = \frac{{\pi {u_F}{\alpha _F}m_F^{a_F}m_E^m\sqrt {1 - v_i^2} }}{{m!\left( {a_F} \right)!2N\lambda _{FS}^{a_F}\lambda _{ES}^m\gamma _0^{a_F}\gamma _E^m}}$, and $h_{F,I}^{asym}\left( x \right) = \frac{{A_{{g_{x,F}}}^{a_F}A_x^{m - 1}{e^{ - \frac{{{m_E}{A_x}}}{{{\gamma _E}{\lambda _{ES}}}}}}}}{{{{\left( {{\alpha _F} - {\alpha _N}x} \right)}^2}}}\left( {\frac{{{m_E}{A_x}}}{{{\gamma _E}{\lambda _{ES}}}} - m} \right)$.

The secrecy diversity order at $F$ in Solution I, $D_{F,I}$, is defined as {\cite{Liu2017}}
\begin{equation}
{D_{F,I}} =  - \mathop {\lim }\limits_{{\gamma _0} \to \infty } \frac{{\log \left[ {SOP_{F,I}^{asym}\left( {{\gamma _0}} \right)} \right]}}{{\log \left( {{\gamma _0}} \right)}}.
\label{sdoFI}
\end{equation}
By substituting (\ref{SOPFI_asym}) into (\ref{sdoFI}), we have $D_{F,I} = 0$.

For $N$, based on (\ref{SOPNi_def}) and (\ref{cdf_hnsI_asym}), and after some algebraic manipulations similar to the proof of Theorem 2 in Appendix C, {the} asymptotic SOP in Solution I is expressed as
\begin{equation}
SOP_{N,I}^{asym} = \left\{ {\begin{array}{*{20}{c}}
{1,}&{{\gamma_{th}} \ge \beta }\\
{\Lambda _{1,I}^{asym} + \Lambda _{2,I}^{asym} + \Lambda _{3,I}^{asym},}&{{\gamma_{th}} < \beta }
\end{array}} \right.,
\label{SOPNI_asym}
\end{equation}
where
\begin{equation}
\Lambda _{1,I}^{asym} \approx \frac{1}{{{{\left[ {\left( {a_N} \right)!} \right]}^{{L_S}}}}}{\left( {\frac{{{m_N}{A_{{\gamma_{th}}}}}}{{{\gamma _0}{\lambda _{NS}}}}} \right)^{{b_N}}}, \nonumber
\end{equation}
\begin{equation}
\Lambda _{2,I}^{asym} \approx \left\{ {\begin{array}{*{20}{c}}
{0,}&{{R_{s,N}} < \eta }\\
{\begin{array}{*{20}{c}}
{\frac{{{{\left( {{m_N}/{\gamma _0}{\lambda _{NS}}} \right)}^{{b_N}}}}}{{{{\left[ {\left( {{a_N}} \right)!} \right]}^{{L_S}}}}}\left( {\gamma _{s,N}^{{b_N}} - A_{{\gamma_{th}}}^{{b_N}}} \right)}\\
{ \times {F_{{Z_{\hat i}}}}\left( {\frac{{{A_{{\gamma_{th}}}}}}{{{\gamma _E}}}} \right) \hspace{2cm}}
\end{array},}&{{R_{s,N}} > \eta }
\end{array}} \right., \nonumber
\end{equation}
\begin{equation}
\Lambda _{3,I}^{asym} \approx B_{N,I}^{asym} + \sum\limits_{m = 0}^{{b_N}} {\sum\limits_{n = 0}^{{a_E} + m -1} {\Psi _{N,I}^{asym}{{\left( {\frac{{{m_E}{A_{{\gamma_{th}}}}}}{{{\gamma _E}{\lambda _{ES}}}}} \right)}^n}{e^{ - \frac{{{m_E}{A_{{\gamma_{th}}}}}}{{{\gamma _E}{\lambda _{ES}}}}}}} }, \nonumber
\end{equation}
with $B_{N,I}^{asym} = \frac{{{{\left( {{m_N}{A_{{\gamma_{th}}}}/{\gamma _0}{\lambda _{NS}}} \right)}^{{b_N}}}}}{{{{\left[ {\left( {a_N} \right)!} \right]}^{{L_S}}}}}\left[ {{F_{{Z_{\hat i}}}}\left( {\frac{{{A_{{\gamma_{th}}}}}}{{{\gamma _E}}}} \right) - 1} \right]$, $\Psi _{N,I}^{asym} = \left( {\begin{array}{*{20}{c}}
{{b_N}}\\
m
\end{array}} \right)\frac{{m_N^{{b_N}}\Gamma \left( {{a_E} + m} \right)\gamma _{s,N}^{{b_N} - m}{{\left( {{2^{{R_{s,N}}}}{\gamma _E}} \right)}^m}\lambda _{ES}^m}}{{n!\Gamma \left( {{a_E}} \right){{\left[ {\left( {{a_N}} \right)!} \right]}^{{L_S}}}{{\left( {{\gamma _0}{\lambda _{NS}}} \right)}^{{b_N}}}m_E^m}}$, and $b_N = a_N{L_S}$.

The secrecy diversity order at $N$ in Solution I, $D_{N,I}$, is given by {\cite{Liu2017}}
\begin{equation}
\begin{split}
{D_{N,I}} &=  - \mathop {\lim }\limits_{{\gamma _0} \to \infty } \frac{{\log \left[ {SOP_{N,I}^{asym}\left( {{\gamma _0}} \right)} \right]}}{{\log \left( {{\gamma _0}} \right)}} \\
&= \left\{ {\begin{array}{*{20}{c}}
{0,}&{{\gamma_{th}} \ge \beta }\\
{{b_N},}&{{\gamma_{th}} < \beta }
\end{array}} \right..
\end{split}
\label{sdoNI}
\end{equation}

Based on (\ref{SOPOi_def}), the overall SOP in Solution I is asymptotically derived as
\begin{equation}
SOP_{O,I}^{asym} = 1 - \left( {1 - SOP_{F,I}^{asym}} \right)\left( {1 - SOP_{N,I}^{asym}} \right),
\label{SOPOI_asym}
\end{equation}
and the achieved overall secrecy diversity order is {\cite{Liu2017}}
\begin{equation}
{D_{O,I}} =  - \mathop {\lim }\limits_{{\gamma _0} \to \infty } \frac{{\log \left[ {SOP_{O,I}^{asym}\left( {{\gamma _0}} \right)} \right]}}{{\log \left( {{\gamma _0}} \right)}} = 0.
\end{equation}

{From the asymptotic and secrecy diversity order results}, we provide some useful insights through the following remarks.

\begin{remark}
 In case of ${\gamma_{th}} < \beta$, the secrecy diversity order of user $N$ is $m_N L_S L_N$. This reveals that Solution I can offer the full secrecy diversity order. Moreover, the increase in $m_N$, $L_S$, and $L_N$ can help improve the secrecy performance of $N$. {However, for the} case of ${\gamma_{th}} \ge \beta$, the zero secrecy diversity order for user $N$ is obtained and hence its secrecy performance is saturated as $\gamma_0 \to \infty$. {In other words, the secrecy performance of $N$ does not depend on the system parameters (i.e., $L_S$, $L_N$, and $m_N$) when $\gamma_0 \to \infty$.}
\end{remark}

\begin{remark}
The secrecy diversity order is $0$ for both user $F$ and the overall system in solution I. {In other words, both user $F$ and the overall system reach a secrecy performance floor in the high $\gamma_0$ regime}. {Similar to Remark 2, this also indicates that the secrecy performance of user $F$ and the overall system are not influenced by $L_S$, $L_N$, $L_F$, $m_N$, and $m_F$ in case of $\gamma_0 \to \infty$.} This will be further clarified in section \ref{Num}.
\end{remark}

\subsection{Secrecy Performance Analysis of Solution II}
\subsubsection{{Secrecy Outage Probability Analysis}}
In this case, the SOP at $F$ and $N$ are derived through Theorems 3 and 4 as follows.

\begin{theorem}
Under Nakagami-$m$ fading, the SOP of $F$ with Solution II is approximately expressed as
\begin{equation}
SOP_{F,II} \approx 1 + \sum\limits_{p,{\Delta _F},m,i}^ \sim  {\Psi _{F,II}^{(1)}{h_{F,II}}\left[ {\frac{{\left( {{v_i} + 1} \right){u_F}}}{2}} \right]},
\label{SOPFII_final}
\end{equation}
where $\sum\limits_{p,{\Delta _F},m,i}^ \sim  {}  = \sum\limits_{p = 1}^{{L_S}} {\sum\limits_{{\Delta _F} = p} {\sum\limits_{m = 0}^{{a_E}} {\sum\limits_{i = 0}^N {} } } }$, $\Psi _{F,II}^{(1)} = \frac{{\pi {u_F}{\Phi _F}{\alpha _F}m_E^m\sqrt {1 - v_i^2} }}{{m!2N\gamma _0^{{\varphi _F}}\gamma _E^m\lambda _{ES}^m}}$, and ${h_{F,II}}\left( x \right) = \frac{{A_{{g_{x,F}}}^{{\varphi _F}}A_x^{m - 1}}}{{{{\left( {{\alpha _F} - {\alpha _N}x} \right)}^2}}}{e^{ - \frac{{p{m_F}{A_{g_{x,F}}}}}{{{\gamma _0}{\lambda _{FS}}}} - \frac{{{m_E}{A_x}}}{{{\gamma _E}{\lambda _{ES}}}}}}\left( {\frac{{m_E}A_x}{{{\gamma _E}{\lambda _{ES}}}} - m} \right)$.
\end{theorem}

\begin{IEEEproof}
The proof of this theorem is similar to the proof of Theorem 1 in Appendix B, in which {Proposition} 2 is used instead of {Proposition} 1.
\end{IEEEproof}

\begin{theorem}
Under Nakagami-$m$ fading, the SOP of $N$ in the case of Solution II is given by
\begin{equation}
SOP_{N,II} = \left\{ {\begin{array}{*{20}{c}}
{1,}&{{\gamma_{th}} \ge \beta }\\
{{\Lambda _{1,II}} + {\Lambda _{2,II}} + {\Lambda _{3,II}},}&{{\gamma_{th}} < \beta }
\end{array}} \right.,
\label{SOPNII_final}
\end{equation}
where
\begin{equation}
\Lambda _{1,II} = {F_{{Y_{\hat i}},II}}\left( {\frac{{{A_{{\gamma_{th}}}}}}{{{\gamma _0}}}} \right), \nonumber
\end{equation}
\begin{small}\begin{equation}
{\Lambda _{2,II}} = \left\{ {\begin{array}{*{20}{c}}
{0,}&{{R_{s,N}} < \eta }\\
{\begin{array}{*{20}{c}}
{\left[ {{F_{{Y_{\hat i}},II}}\left( {\frac{{{\gamma _{s,N}}}}{{{\gamma _0}}}} \right) - {F_{{Y_{\hat i}},II}}\left( {\frac{{{A_{{\gamma_{th}}}}}}{{{\gamma _0}}}} \right)} \right]}\\
{ \times {F_{{Z_{\hat i}}}}\left( {\frac{{{A_{{\gamma_{th}}}}}}{{{\gamma _E}}}} \right) \hspace{2.7cm}}
\end{array},}&{{R_{s,N}} > \eta }
\end{array}} \right., \nonumber
\end{equation}
\end{small}and
\begin{equation}
{\Lambda _{3,II}} = B_{N,II}^{(1)} - \sum\limits_{m,n,p}^{ \sim}  {{\Psi _{N,II}}{{\left[ {\frac{{{A_{{\gamma_{th}}}}B_{N,II}^{(2)}}}{{{\gamma _E}}}} \right]}^n}{e^{ - \frac{{{A_{{\gamma_{th}}}}B_{N,II}^{(2)}}}{{{\gamma _E}}}}}}, \nonumber
\end{equation}
with $B_{N,II}^{(1)} = \left[ {1 - {F_{{Y_{\hat i}},II}}\left( {\frac{{{A_{{\gamma_{th}}}}}}{{{\gamma _0}}}} \right)} \right]\left[ {1 - {F_{{Z_{\hat i}}}}\left( {\frac{{{A_{{\gamma_{th}}}}}}{{{\gamma _E}}}} \right)} \right]$, $B_{N,II}^{(2)} = \frac{{{m_E}}}{{{\lambda _{ES}}}} + \frac{{{m_N}{\gamma _E}{2^{{R_{s,N}}}}}}{{{\gamma _0}{\lambda _{NS}}}}$, $\sum\limits_{m,n,p}^ \sim  {}  = \sum\limits_{m = 0}^{{a_N -1}} {\sum\limits_{n = 0}^m {\sum\limits_{p = 0}^{{a_E} + n -1}}}$, $F_{{Y_{\hat i}},II}\left( x \right)$ is defined in (\ref{cdf_hnsII_final}), and ${\Psi _{N,II}} = \left( {\begin{array}{*{20}{c}}
m\\
n
\end{array}} \right)\frac{{m_N^mm_E^{{a_E}}\Gamma \left( {{a_E} + n} \right)\gamma _{s,N}^{m - n}{{\left( {{2^{{R_{s,N}}}}{\gamma _E}} \right)}^n}{e^{ - \frac{{{m_N}{\gamma _{s,N}}}}{{{\gamma _0}{\lambda _{NS}}}}}}}}{{m!p!\Gamma \left( {{a_E}} \right)\lambda _{ES}^{{a_E}}\lambda _{NS}^m\gamma _0^m{{\left[ {B_{N,II}^{(2)}} \right]}^{{a_E} + n}}}}$.
\end{theorem}

\begin{IEEEproof}
The proof of this theorem is similar to the proof of Theorem 2 in Appendix C, in which $F_{{Y_{\hat i}},II}\left( x \right)$ in (\ref{cdf_hnsII_final}) is used instead of $F_{{Y_{\hat i}},I}\left( x \right)$ in (\ref{cdf_hnsI_final}).
\end{IEEEproof}

Based on (\ref{SOPFII_final}) and (\ref{SOPNII_final}), the overall SOP of the system in the case of Solution II is given by
\begin{equation}
SOP_{O,II} = 1 - \left( 1 - SOP_{F,II} \right)\left( 1 - SOP_{N,II} \right).
\label{SOPOII_def}
\end{equation}

\subsubsection{Asymptotic Secrecy Outage Probability Analysis \label{Asymp2}}

Following the similar calculation steps as in \ref{Asymp1}, with Solution II, the asymptotic SOP of user $F$ is approximated as
\begin{equation}
\begin{split}
SOP_{F,II}^{asym} &\approx \sum\limits_{m = 0}^{{a_E -1}} {\sum\limits_{i = 0}^N {\Psi _{F,II}^{asym}h_{F,II}^{asym}\left[ {\frac{{\left( {{v_i} + 1} \right){u_F}}}{2}} \right]}} \\
& \quad + \sum\limits_{m = 0}^{{a_E -1}} {\Psi _{F,E}},
\end{split}
\label{SOPFII_asym}
\end{equation}
where $\Psi _{F,II}^{asym} = \frac{{\pi {u_F}{\alpha _F}\sqrt {1 - v_i^2} }}{{m!{{\left[ {\left( {a_F} \right)!} \right]}^{{L_S}}}2N}}{\left( {\frac{{{m_F}}}{{{\gamma _0}{\lambda _{FS}}}}} \right)^{{b_F}}}{\left( {\frac{{{m_E}}}{{{\gamma _E}{\lambda _{ES}}}}} \right)^m}$, $h_{F,II}^{asym}\left( x \right) = \left( {\frac{{{m_E}{A_x}}}{{{\gamma _E}{\lambda _{ES}}}} - m} \right) \frac{{A_{{g_{x,F}}}^{{b_F}}A_x^{m - 1}{e^{ - \frac{{{m_E}{A_x}}}{{{\gamma _E}{\lambda _{ES}}}}}}}}{{{{\left( {{\alpha _F} - {\alpha _N}x} \right)}^2}}}$, and $b_F = a_F L_S$.

From (\ref{SOPFII_asym}), we conclude that {the} secrecy diversity order at $F$ in this case is $D_{F,II} = 0$.

Next, the asymptotic SOP at $N$ is given by
\begin{equation}
SOP_{N,II}^{asym} = \left\{ {\begin{array}{*{20}{c}}
{1,}&{{\gamma_{th}} \ge \beta }\\
{\Lambda _{1,II}^{asym} + \Lambda _{2,II}^{asym} + \Lambda _{3,II}^{asym},}&{{\gamma_{th}} < \beta }
\end{array}} \right.,
\label{SOPNII_asym}
\end{equation}
where 
\begin{equation}
\Lambda _{1,II}^{asym} \approx \frac{1}{{\left( {a_N} \right)!}}{\left( {\frac{{{m_N}{A_{{\gamma_{th}}}}}}{{{\gamma _0}{\lambda _{NS}}}}} \right)^{a_N}}, \nonumber
\end{equation}
\begin{small}
\begin{equation}
\Lambda _{2,II}^{asym} \approx \left\{ {\begin{array}{*{20}{c}}
{0,}&{{R_{s,N}} < \eta }\\
{\begin{array}{*{20}{c}}
{\frac{{{{\left( {{m_N}/{\gamma _0}{\lambda _{NS}}} \right)}^{{a_N}}}}}{{\left( {{a_N}} \right)!}}{F_{{Z_{\hat i}}}}\left( {\frac{{{A_{{\gamma_{th}}}}}}{{{\gamma _E}}}} \right)}\\
{ \times \left( {\gamma _{s,N}^{{a_N}} - A_{{\gamma_{th}}}^{{a_N}}} \right) \hspace{1.5cm}}
\end{array},}&{{R_{s,N}} > \eta }
\end{array}} \right., \nonumber
\end{equation}
\end{small}
\begin{equation}
\Lambda _{3,II}^{asym} \approx B_{N,II}^{asym} + \sum\limits_{m = 0}^{a_N} {\sum\limits_{n = 0}^{{a_E} + m -1} {\Psi _{N,II}^{asym}{{\left[ {\frac{{{m_E}{A_{{\gamma_{th}}}}}}{{{\gamma _E}{\lambda _{ES}}}}} \right]}^n}{e^{ - \frac{{{m_E}{A_{{\gamma_{th}}}}}}{{{\gamma _E}{\lambda _{ES}}}}}}} }, \nonumber
\end{equation}
with $B_{N,II}^{asym} = \frac{{{{\left( {{m_N}{A_{{\gamma_{th}}}}/{\gamma _0}{\lambda _{NS}}} \right)}^{a_N}}}}{{\left( {a_N} \right)!}}\left[ {{F_{{Z_{\hat i}}}}\left( {\frac{{{A_{{\gamma_{th}}}}}}{{{\gamma _E}}}} \right) - 1} \right]$, and $\Psi _{N,II}^{asym} = \left( {\begin{array}{*{20}{c}}
{{a_N}}\\
m
\end{array}} \right)\frac{{m_N^{{a_N}}\Gamma \left( {{a_E} + m} \right)\gamma _{s,N}^{{a_N} - m}{{\left( {{2^{{R_{s,N}}}}{\gamma _E}} \right)}^m}\lambda _{ES}^m}}{{n!\left( {{a_N}} \right)!\Gamma \left( {{a_E}} \right){{\left( {{\gamma _0}{\lambda _{NS}}} \right)}^{{a_N}}}m_E^m}}$.

It is shown from (\ref{SOPNII_asym}) that the secrecy diversity order at user $N$ in the case of Solution II {is obtained as}
\begin{equation}
{D_{N,II}}= \left\{ {\begin{array}{*{20}{c}}
{0,}&{{\gamma_{th}} \ge \beta }\\
{a_N,}&{{\gamma_{th}} < \beta }
\end{array}} \right..
\label{sdoNII}
\end{equation}

From (\ref{SOPFII_asym}) and (\ref{SOPNII_asym}), the asymptotic overall SOP {is given by}
\begin{equation}
SOP_{O,II}^{asym} = 1 - \left( {1 - SOP_{F,II}^{asym}} \right)\left( {1 - SOP_{N,II}^{asym}} \right),
\label{SOPOII_asym}
\end{equation}
and the achieved overall secrecy diversity order in this case is $D_{O,II} = 0$.

\begin{remark}
It is worth noting that Solution II provides the secrecy diversity order $m_N L_N$ (if ${\gamma_{th}} < \beta$) and $0$ (if ${\gamma_{th}} \ge \beta$) for $N$. Thus, in this case, the secrecy performance of $N$ does not depend on $L_S$. Also, it can be improved by increasing $m_N$ and $L_N$. Solution II also {shows} the zero secrecy diversity order for both $F$ and the overall system similar to Solution I. 
\end{remark}

\section{Numerical Results and Discussions \label{Num}}
In this section, numerical results, in terms of the SOP, are provided to evaluate the secrecy performance of our solutions for designing {a} secure MIMO NOMA network. 

\begin{figure}[!t]
	\centering
	\includegraphics[scale = 0.35]{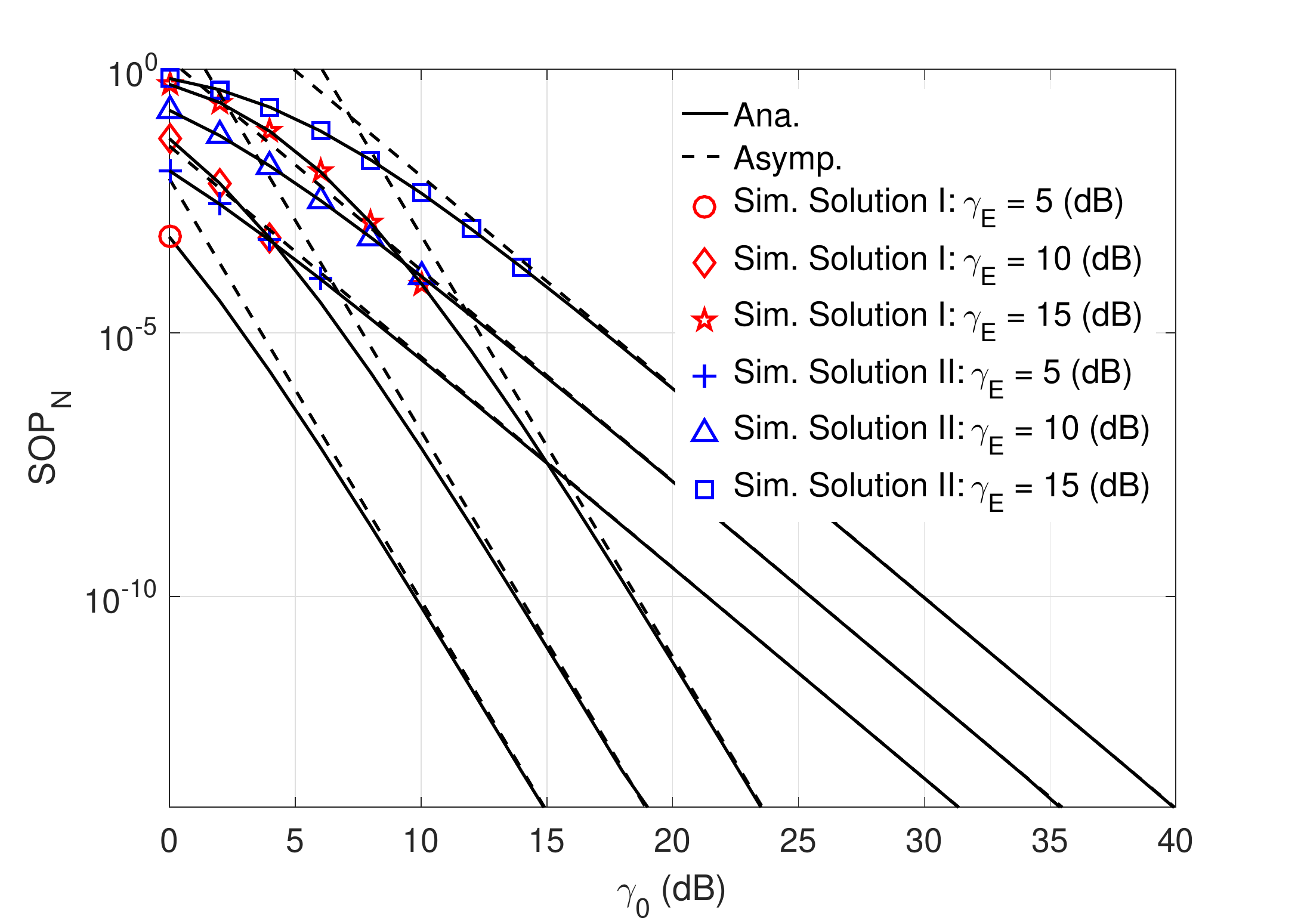}
	\caption{$SOP_N$ v.s. $\gamma_0$ with different values of $\gamma_E${,} where {$m_F = m_N = m_E = 2$}, $L_S = L_N = L_E = 2$, $\alpha_F = 0.6$, {and} $\alpha_N = 0.4$.} 
	\label{sopn_g0e}
\end{figure}

\begin{figure}[!t]
	\centering
	\includegraphics[scale = 0.35]{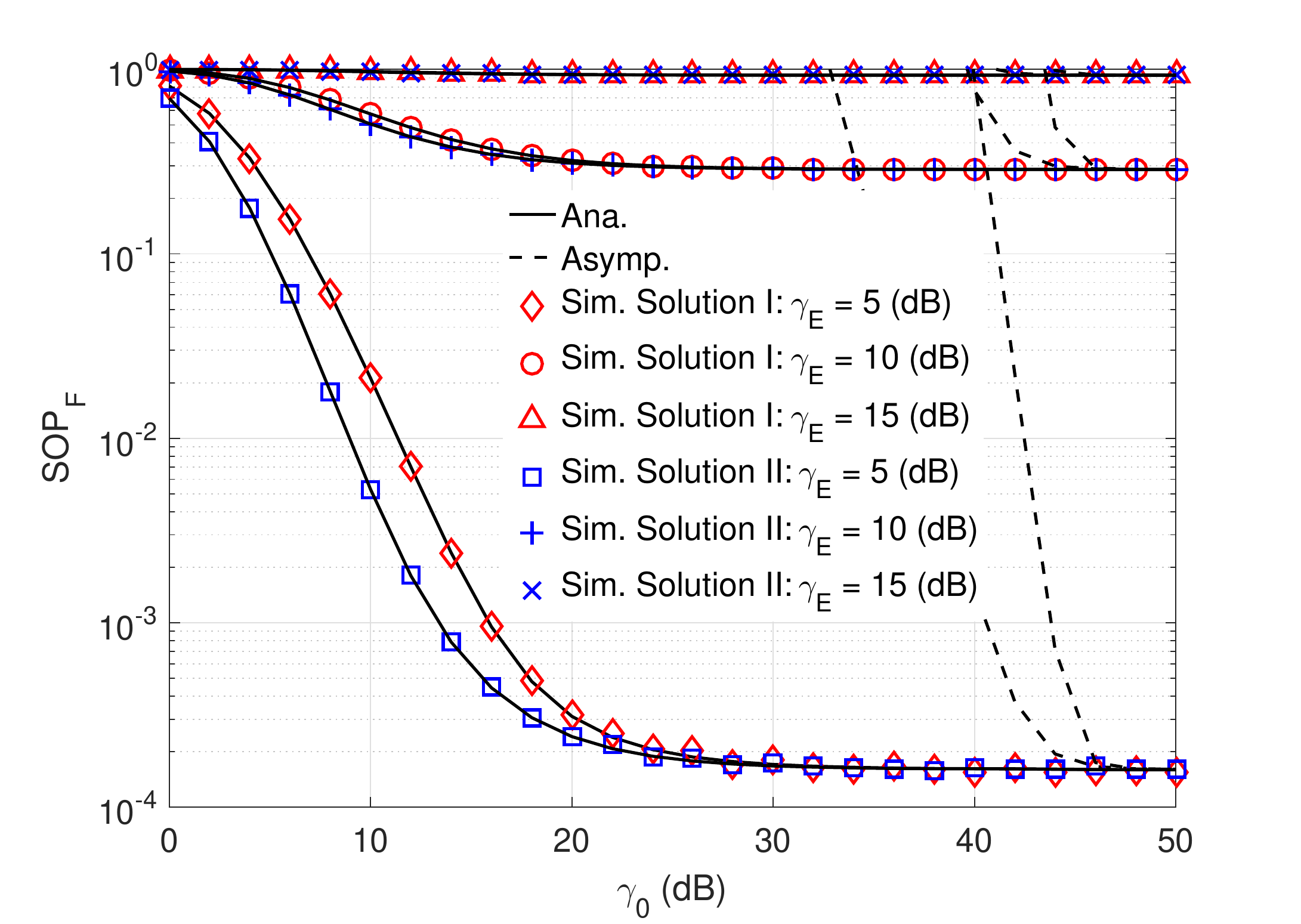}
	\caption{$SOP_F$ v.s. $\gamma_0$ with different values of $\gamma_E$, where {$m_F = m_N = m_E = 2$}, $L_S = L_F = L_E = 2$, $\alpha_F = 0.6$, {and} $\alpha_N = 0.4$.} 
	\label{sopf_g0e}
\end{figure}

Specifically, the SOP of $F$ (i.e., $SOP_F$), $N$ (i.e., $SOP_N$), and the overall system (i.e., $SOP_O$) are investigated {for} two different TAS solutions (i.e., Solutions I and II). Without loss of generality, it is assumed that the distance from $S$ to $F$ is set to unity and $N$ is located {in between them}. Indeed, this setting has been widely used in the literature \cite{Che2015, Fan2016}. In more details, the coordinates of $S$, $F$, $N$, and $E$ are $(0{,} 0.5)$, $(1{,} 0.5)$, $(0.5{,} 0.5)$, and $(3{,} 0)$, respectively. Furthermore, the predetermined simulation parameters are set as follows: the target data rate and the secrecy rate thresholds $R_F = R_{s,F} = R_{s,N} = 0.5$ (bps/Hz), the path-loss exponents $\theta_{FS} = \theta_{NS} = \theta_{ES} = 2$. {For the complexity-accuracy tradeoff parameter $N$, we can define an its specific value to achieve the expected analytical results \cite{Liu2017, Liu20161}. In this paper, we use $N = 100$ for our analysis. Note that in the legends of all obtained results, we use the following abbreviations: Ana. stands for the analytical results, Asymp. denotes the asymptotic results, and Sim. indicates the simulation results.}

%\textcolor{red}{Note that in the legends of all obtained results, we use the following abbreviations: Ana. stands for the analytical results achieved from \eqref{SOPFI_final} (for user $F$), \eqref{SOPNI_final} (for user $N$), \eqref{SOPOI_def} (for overall system) with Solution I and from \eqref{SOPFII_final} (for user $F$), \eqref{SOPNII_final} (for user $N$), \eqref{SOPOII_def} (for overall system) with Solution II; Asymp. denotes the asymptotic results given by \eqref{SOPFI_asym} (for user $F$), \eqref{SOPNI_asym} (for user $N$), \eqref{SOPOI_asym} (for overall system) with Solution I and \eqref{SOPFII_asym} (for user $F$), \eqref{SOPNII_asym} (for user $N$), \eqref{SOPOII_asym} (for overall system) with Solution II; Sim. indicates the simulation results.}

\begin{figure}[!t]
	\centering
	\includegraphics[scale = 0.35]{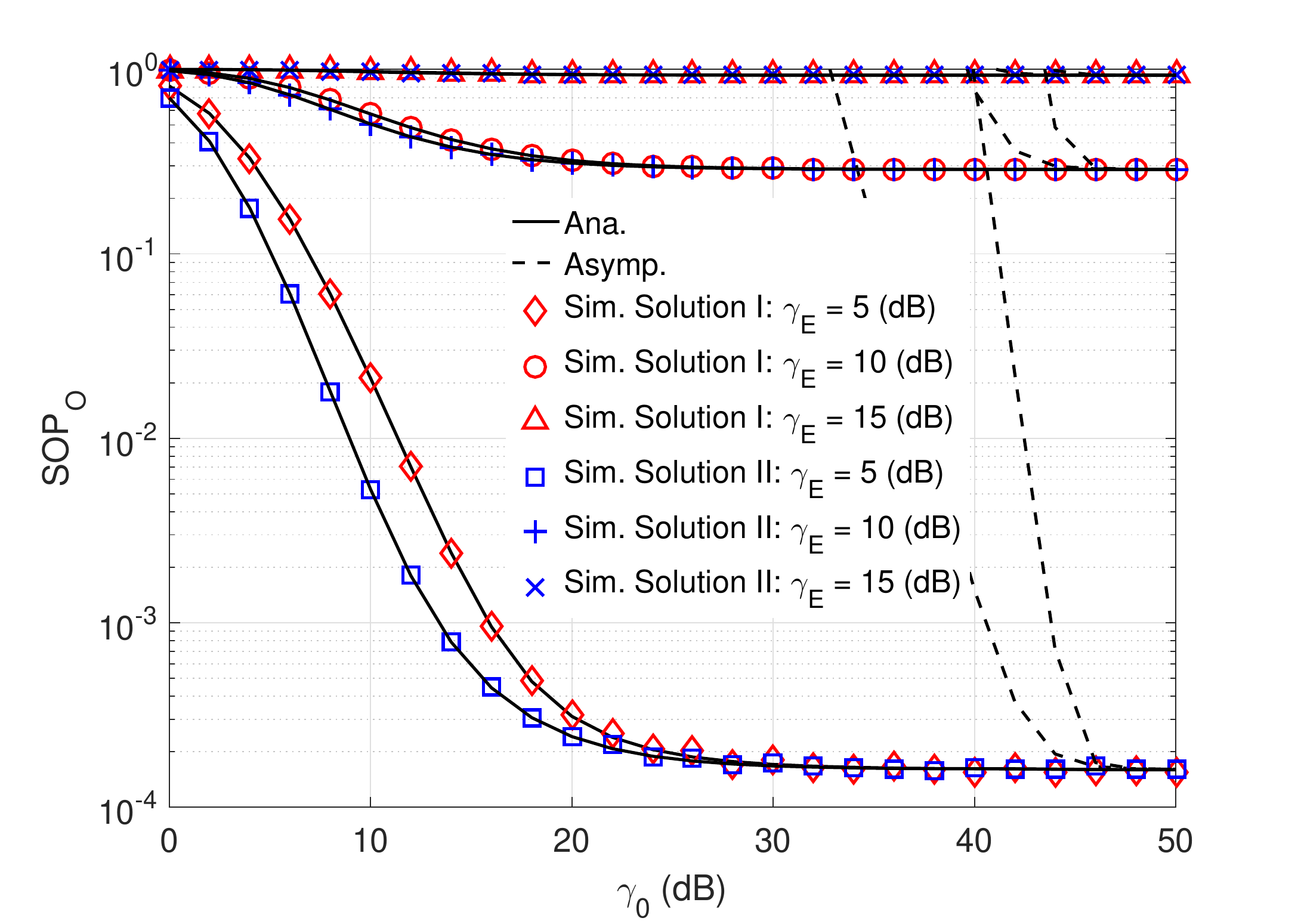}
	\caption{$SOP_O$ v.s. $\gamma_0$ with the different values of $\gamma_E$, where {$m_F = m_N = m_E = 2$}, $L_S = L_F = L_N = L_E = 2$, $\alpha_F = 0.6$, {and} $\alpha_N = 0.4$.} 
	\label{sopo_g0e}
\end{figure}

In Figs. \ref{sopn_g0e}, \ref{sopf_g0e}, and \ref{sopo_g0e}, we examine $SOP_N$, $SOP_F$, and $SOP_O$ as a function of average transmit SNR for LUs ($\gamma_0$) with different values of $\gamma_E$. It is observed that the secrecy performance evaluated at $N$, $F$ and for the overall system are significantly improved (i.e., $SOP_N$, $SOP_F$, and $SOP_O$ decrease) when the higher values of $\gamma_0$ and the smaller values of $\gamma_E$ are considered. Particularly, one can see that $SOP_F$ and $SOP_O$ remain constant in the high $\gamma_0$ regime, i.e., $\gamma_0 \to \infty$. In other words, the secrecy performance obtained at $F$ and the secrecy performance of the overall system are saturated. Then, this phenomenon confirms the calculation of the zero secrecy diversity order for $F$ and the overall system in both Solution I and Solution II, as indicated in \ref{Asymp1} and \ref{Asymp2}.

\begin{figure}[!t]
	\centering
	\includegraphics[scale = 0.35]{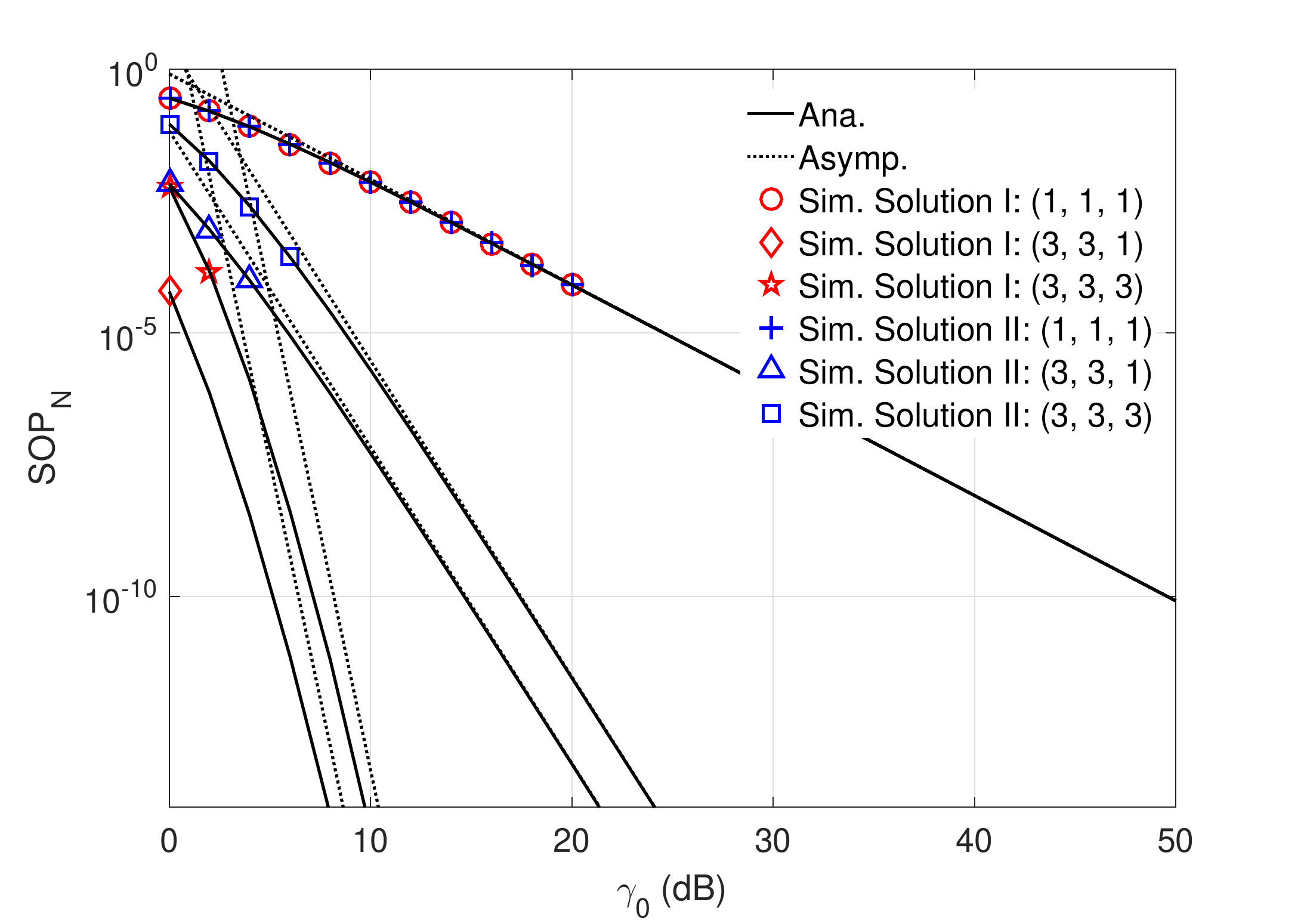}
	\caption{$SOP_N$ v.s. $\gamma_0$ with different values of ($L_S{,} L_N{,} L_E$){,} where {$m_F = m_N = m_E = 2$}, $\alpha_F = 0.6$, $\alpha_N = 0.4$, {and} $\gamma_E = 10$ (dB).} 
	\label{sopn_g0L}
\end{figure}

\begin{figure}[!t]
	\centering
	\includegraphics[scale = 0.35]{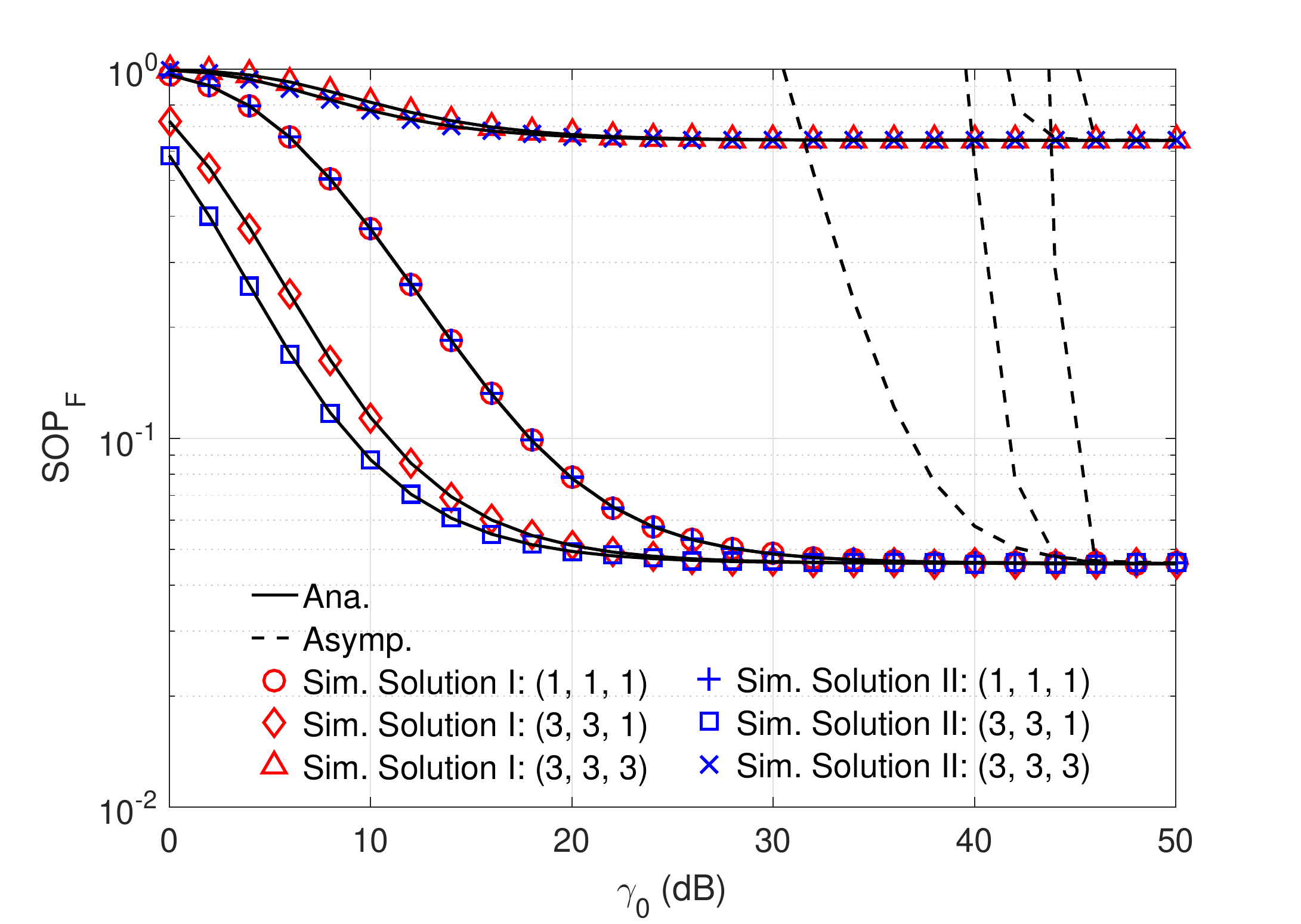}
	\caption{$SOP_F$ v.s. $\gamma_0$ with different values of ($L_S{,} L_F{,} L_E$), where {$m_F = m_N = m_E = 2$}, $\alpha_F = 0.6$, $\alpha_N = 0.4$, {and} $\gamma_E = 10$ (dB).} 
	\label{sopf_g0L}
\end{figure}

\begin{figure}[!t]
	\centering
	\includegraphics[scale = 0.35]{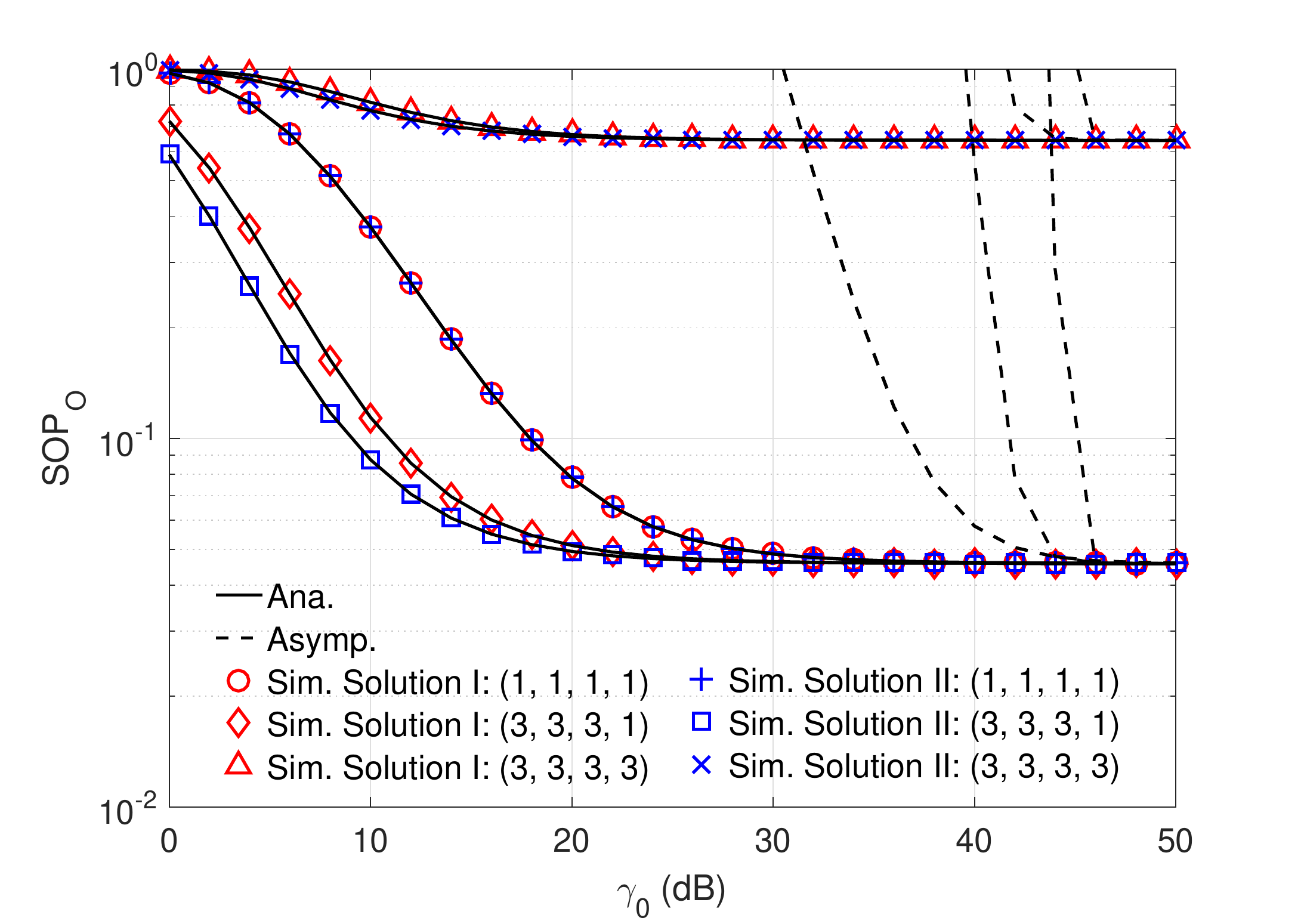}
	\caption{$SOP_O$ v.s. $\gamma_0$ with the different values of $(L_S{,} L_F{,} L_N{,} L_E)$, where {$m_F = m_N = m_E = 2$}, $\alpha_F = 0.6$, $\alpha_N = 0.4$, {and} $\gamma_E = 10$ (dB).} 
	\label{sopo_g0L}
\end{figure}

In order to study how using multiple antennas impacts the secrecy performance, we investigate the variations of $SOP_N$, $SOP_F$, and $SOP_O$ with respect to the different values of the number of antennas at $S$, $F$, $N$, and $E$, denoted by $(L_S{,} L_F{,} L_N {,} L_E)$, as illustrated in Figs. \ref{sopn_g0L}, \ref{sopf_g0L}, and \ref{sopo_g0L}, respectively. Particularly, in the case of $(L_S{,} L_F{,} L_N{,} L_E) = (1{,} 1{,} 1{,} 1)$, Solutions I and II yield the same curves. In general, it is visible that employing MIMO at the LUs improves the secrecy performance significantly. In addition, the more antennas the eavesdropper has, the worse {the} security performance the system is. More specific discussions regarding these figures are following.

As observed in Fig. \ref{sopn_g0L}, { decreasing} in $L_E$ and {increasing} in $L_N$ and $L_S$ {positively effect the secrecy performance} evaluated at $N$. Furthermore, as shown in \ref{Asymp1} and \ref{Asymp2}, the secrecy diversity order at $N$ are $m_N L_N L_S$ and $m_N L_N$ {for the case} of Solution I  and Solution II (${\gamma_{th}} < \beta$), respectively. In other words, considering Solution I, a {significant} improvement of the secrecy performance can be obtained by increasing $m_N$, $L_N$, and $L_S$. However, accounting for Solution II, the improvement can only occur {by increasing} $m_N$ and $L_N$.

In Fig. \ref{sopf_g0L}, {the} $SOP_F$ decreases when $L_F$ and $L_S$ increase as well as $L_E$ decreases. However, {as mentioned earlier}, in the high $\gamma_0$ regime, the secrecy performance at $F$ achieves the saturation which does not depend on the number of antennas (i.e., $L_F$ and $L_S$). {This} phenomenon {is also observed} when considering the effect of $(L_S{,} L_F{,} L_N{,} L_E)$ on $SOP_O$ in Fig. \ref{sopo_g0L}. Specifically, the overall secrecy performance can be improved {when} $L_S$, $L_F$, and $L_N$ scale up or $L_E$ scales down in the low and medium range of $\gamma_0$. Moreover, in the high $\gamma_0$ regime, $L_S$, $L_F$, and $L_N$ have no impact on $SOP_O$ and then the overall system reaches a secrecy performance floor.

{In NOMA systems}, the secrecy performance depends on the power allocation coefficients, i.e., $\alpha_F$ and $\alpha_N$. Fig. 8 is provided to clarify the effect of $\alpha_F$ and $\alpha_N$ on the secrecy performance of the overall system, where $\alpha_F > \alpha_N > 0$ and $\alpha_N = 1 - \alpha_F$. {From this figure, one can see that there is a pair of optimal values of  $\alpha_F$ and $\alpha_N$, denoted by $\left( \alpha_F^*, \alpha_N^* \right)$, that maximizes the overall secrecy performance. This can be explained by the fact that when $\alpha_F$ increases within $\alpha_F < \alpha_F^*$, the secrecy performance of user $F$ is improved significantly. This leads to the improvement of the overall secrecy performance. However, when $\alpha_F$ increases within $\alpha_F > \alpha_F^*$, the secrecy performance of user $N$ decreases considerably due to a significant reduction of $\alpha_N$. Hence, the loss of the overall secrecy performance can be observed. Furthermore, this figure shows that Solution II outperforms Solution I in terms of the overall secrecy performance. Therefore, $\alpha_F$ in Solution I should be higher than in Solution II to obtain a better secrecy performance. In this figure, $\left( \alpha_F^*, \alpha_N^* \right)$ is roughly equal to $\left( 0.98, 0.02 \right)$ for Solution I and $\left( 0.92, 0.08 \right)$ for Solution II.}

\begin{figure}[!t]
\centering
\includegraphics[scale = 0.35]{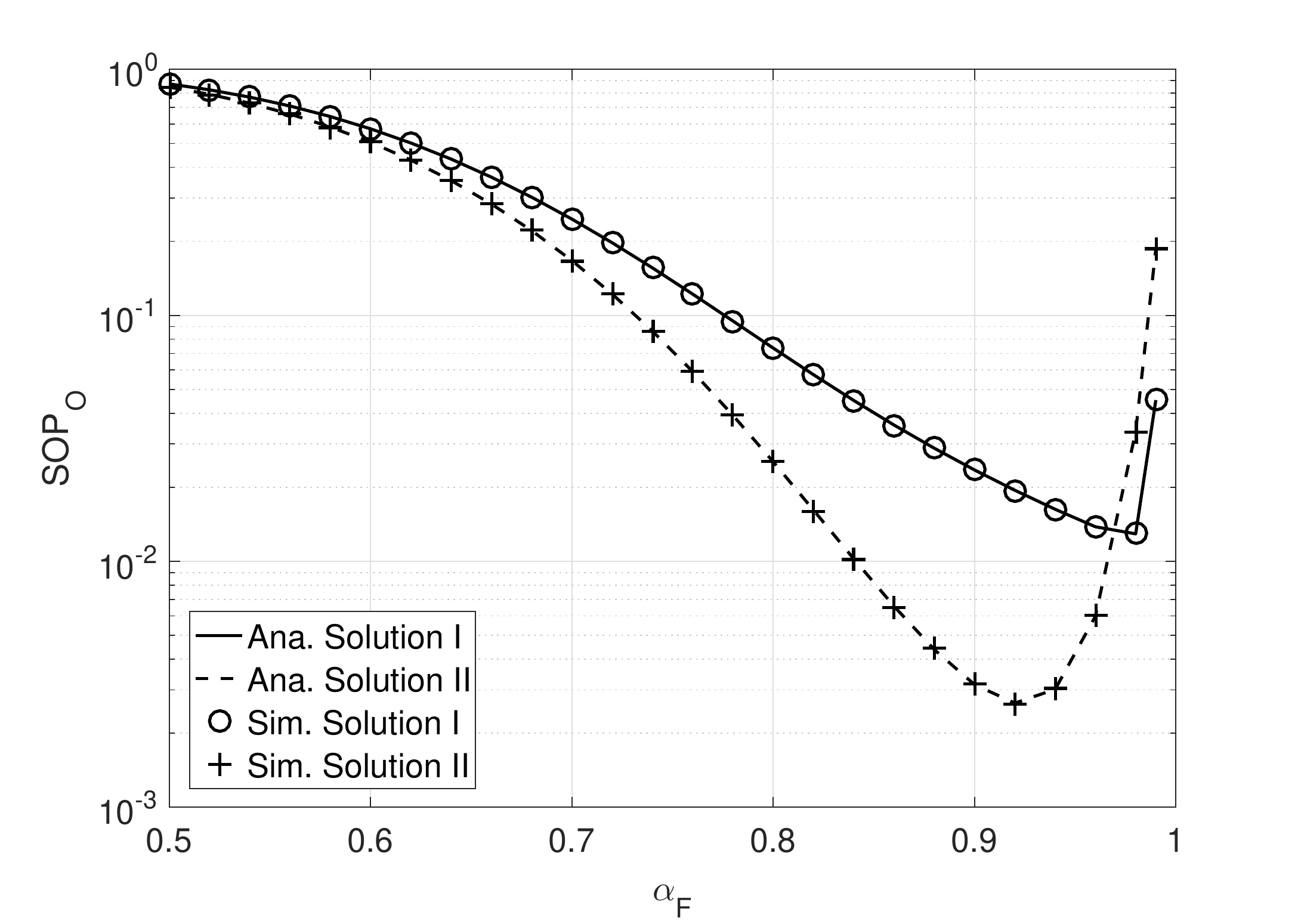}
\caption{$SOP_O$ v.s. $\alpha_F$, where {$m_F = m_N = m_E = 2$}, $L_S = L_F = L_N = L_E = 2$ {and} $\gamma_0 = \gamma_E = 10$ (dB).} 
\label{sopo_alphaf}
\end{figure}

\begin{figure}[!t]
\centering
\includegraphics[scale = 0.35]{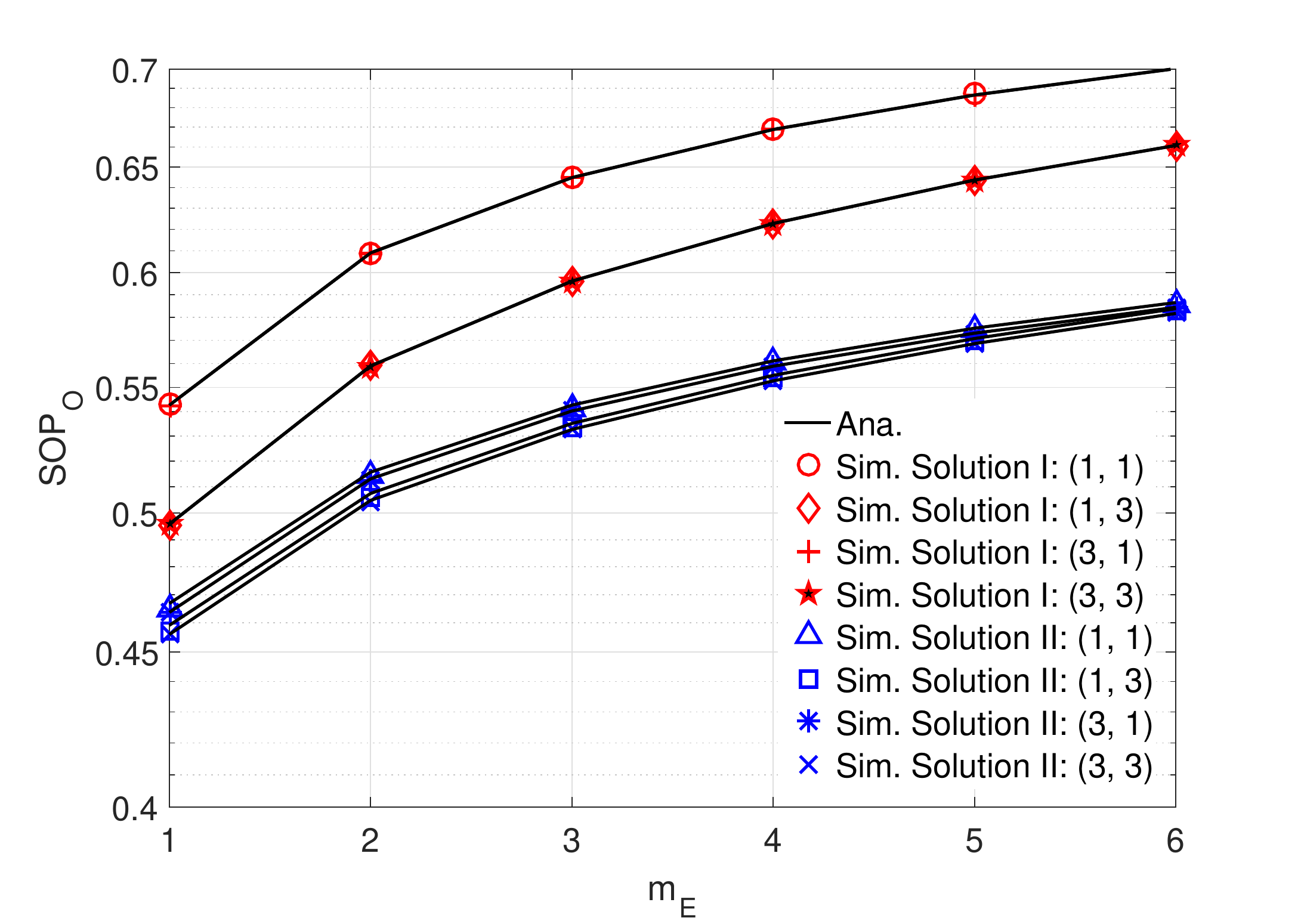}
\caption{{$SOP_O$ v.s. $m_E$ with different values of $\left(m_N, m_F\right)$, where $\alpha_F = 0.6$, $\alpha_N = 0.4$, $L_S = L_F = L_N = L_E = 2$, and $\gamma_0 = \gamma_E = 10$ (dB).}}
\label{sopo_me_mnf}
\end{figure}

{In Fig. \ref{sopo_me_mnf}, the effect of {the} fading parameters, i.e., $m_N$, $m_F$, and $m_E$, on the overall secrecy performance is investigated. {It is recalled that} Nakagami-$m$ fading corresponds to Rayleigh fading as $m = 1$, whereas it approximates Rician fading with parameter $K$ as $m = \left(K + 1\right)^2/\left(2K + 1\right)$. We can see from this figure that the improvement of the secrecy performance is observed with the increases in $m_N$ and $m_F$ and the decrease in $m_E$. This can be explained {by the fact that a} better legitimate channel quality and {a} worse illegitimate channel quality are obtained when increasing $m_N$ and $m_F$ and decreasing $m_E$, respectively. In addition, Fig. \ref{sopo_me_mnf} indicates that $m_F$ has a stronger impact on the overall secrecy performance than $m_N$ does.}

\begin{figure}[!t]
\centering
\includegraphics[scale = 0.35]{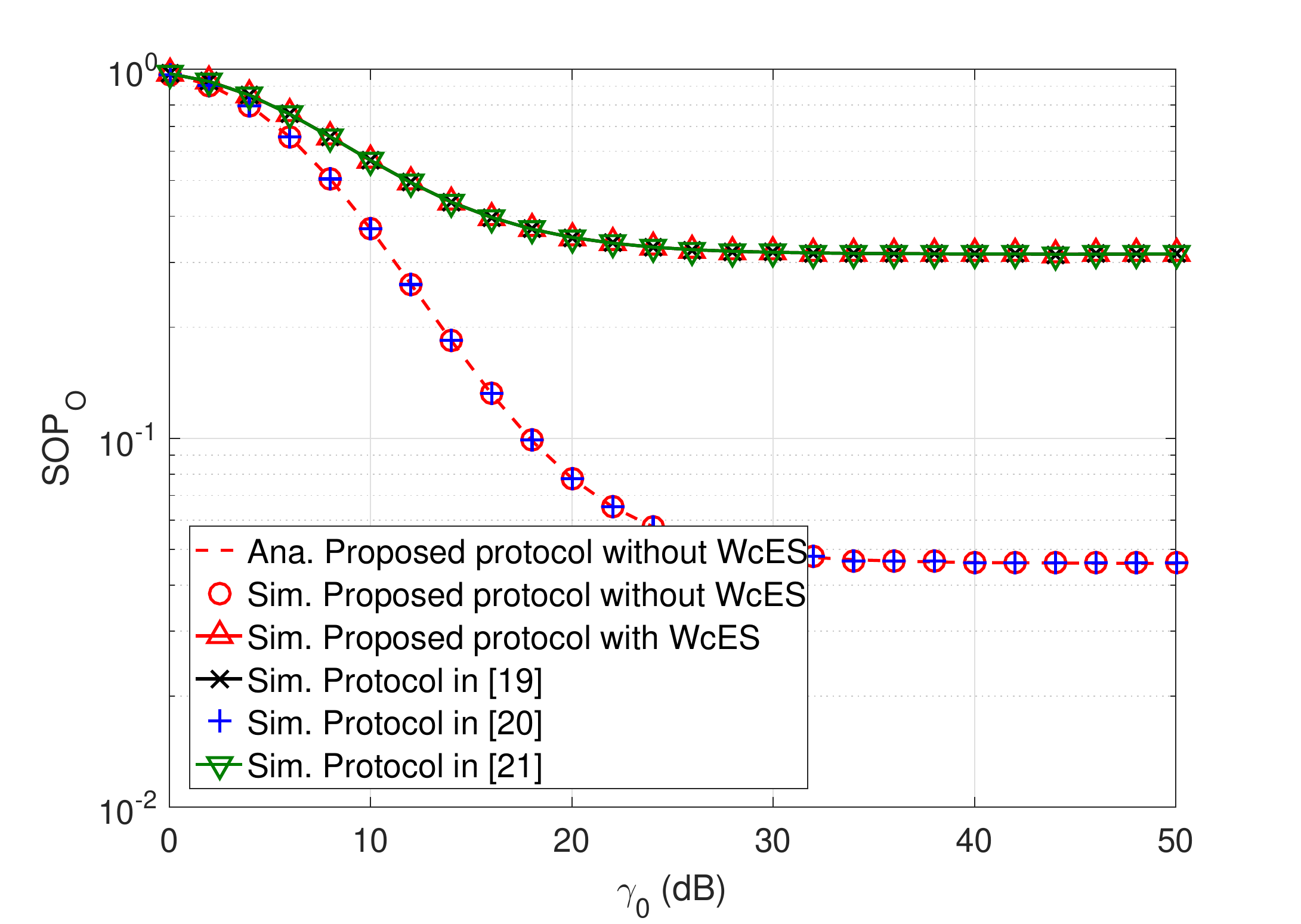}
\caption{{Comparison of secure communication protocols with Solution I, where $m_F = m_N = m_E = 2$, $\alpha_F = 0.6$, $\alpha_N = 0.4$, $\gamma_E = 10$ (dB), $L_S = 2$, and $L_F = L_N = L_E = 1$.}}
\label{sopo_sic}
\end{figure}

{In order to clarify the effectiveness of our proposed protocol, the comparison between our protocol and conventional ones given in \cite{Liu2017, Lei2017, Lv2018}, denoted by $SOP_O$, is presented in Fig. \ref{sopo_sic}. Furthermore, the results regarding the WcES are provided in the figure as a benchmark. 
It can be observed from Fig. \ref{sopo_sic} that in our proposed protocol, the WcES yields a significant decrease in the secrecy performance. This is because the eavesdropper has the ability of powerfully detecting multi-user data without the interference in this case. Comparing with the previous works \cite{Liu2017, Lei2017, Lv2018}, we obtain the following different results:}
{\begin{itemize}
\item {Our protocol gives a much better overall secrecy performance, compared to the protocol in \cite{Liu2017}. This can be explained by the fact that the solution in \cite{Liu2017} overestimates the eavesdropper's multi-user decodability, in which the eavesdropper can decode multi-user data streams without the interference generated by superposed transmit signals. This assumption may be a non-trivial task and hence leads to a significant reduction of the secrecy performance. Furthermore, it can be seen that with WcES, our protocol and the method in \cite{Liu2017} can obtain  similar results.}
\item Fig. \ref{sopo_sic} shows that nothing of significance has changed in the secrecy performance obtained from our protocol and the method in \cite{Lei2017}. However, our protocol brings more practical and general insights. {Specifically, the assumption used in \cite{Liu2017, Lei2017} (i.e., the strong user always successfully decodes the message of the weak user) is a strong one and may be difficult to achieve in realistic scenarios. Whereas, our protocol considers the realistic case where the strong user could successfully or unsuccessfully decode the message of the weak user to analyze the secrecy performance.}
\item Our protocol achieves better secrecy performance than the protocol in \cite{Lv2018} since the solution in \cite{Lv2018} still considered WcES. Note that our protocol can achieve the results similar to the solution in \cite{Lv2018} in case of the WcES.
\end{itemize}}

{Besides, unlike the works \cite{Liu2017, Lei2017, Lv2018} considering MISO systems and Rayleigh fading, our paper investigates the secrecy performance of a more general NOMA system with MIMO and Nakagami-$m$ fading.}

It is noteworthy from all results that the secrecy performance at $N$ is much better than that at $F$ (i.e., $SOP_N$ is much less than $SOP_F$ ). It can be explained by the fact that $S$ $\to$ $N$ link has a better channel gain than $S$ $\to$ $F$ link does since $N$ is closer to $S$ than $F$ does. In addition, one can see that Solution I brings a better secrecy performance for $N$ over Solution II, whereas Solution II offers a higher secrecy performance for $F$ than Solution I does. This can be explained according to the principle of TAS solutions presented in Section \ref{tas}. Taking the overall secrecy performance into account, it can be evaluated that Solution II outperforms Solution I. This implies guaranteeing the secure performance at $F$ has a stronger impact on the overall secrecy performance than that at $N$.

\section{Conclusion}
In this work, we considerd a two-user MIMO NOMA network in the presence of an eavesdropper over Nakagami-$m$ channels. Accordingly, we proposed two TAS solutions to design the secure communication protocol for the network. To validate the solutions, we evaluated the secrecy performance of the system by deriving the exact and asymptotic closed-form expressions of the SOP at the LUs and the SOP of the overall system. Numerical results demonstrated that increasing the number of antennas at the source and the LUs enhances the overall secrecy performance in the low and medium ranges of the average transmit SNR values ($\gamma_0$), however, this has no impact on the performance when $\gamma_0 \to \infty$. In addition, properly choosing $\alpha_F$ and $\alpha_N$ can yield a better overall secrecy performance.  Furthermore, our proposed protocol brings not only an improved secrecy performance but also more practical and general insights, compared with previous protocols. Interestingly, Solution II outperforms Solution I regarding the secrecy performance even {though} they achieve the same zero overall secrecy diversity order.

\appendices
\section{Proof of Equation (\ref{SOPNi_def})}
We rewrite $\Theta_1$, $\Theta_2$, and $\Theta_3$ in (\ref{SOPNi_def}) as follows:
\begin{equation}
\begin{split}
{\Theta _1} &= \Pr \left\{ {\left( {{\alpha _F} - {\alpha _N}{\gamma_{th}}} \right){\gamma _0}{Y_i} < {\gamma_{th}}} \right\} \\
&= \left\{ {\begin{array}{*{20}{c}}
{1,}&{{\gamma_{th}} \ge \beta }\\
{{F_{{Y_i}}}\left( {\frac{{{A_{{\gamma_{th}}}}}}{{{\gamma _0}}}} \right) = {\Lambda _1},}&{{\gamma_{th}} < \beta }
\end{array}} \right.,
\end{split}
\label{Theta1_cal}
\end{equation}
\begin{equation}
\begin{split}
{\Theta _2} &= \Pr \left\{ {\left( {{\alpha _F} - {\alpha _N}{\gamma_{th}}} \right){\gamma _0}{Y_i} \ge {\gamma_{th}},} \right. \\
&\hspace{1cm} \left. {\left( {{\alpha _F} - {\alpha _N}{\gamma_{th}}} \right){\gamma _E}{Z_i} < {\gamma_{th}},{Y_i} < {\gamma _{s,N}}/{\gamma _0}} \right\} \\
& =\left\{ {\begin{array}{*{20}{c}}
{0,}&{{\gamma_{th}} \ge \beta }\\
{\underbrace {\Pr \left\{ {{Y_i} \ge \frac{{{A_{{\gamma_{th}}}}}}{{{\gamma _0}}},{Z_i} < \frac{{{A_{{\gamma_{th}}}}}}{{{\gamma _E}}},{Y_i} < \frac{{{\gamma _{s,N}}}}{{{\gamma _0}}}} \right\}}_{{\Lambda _2}},}&{{\gamma_{th}} < \beta }
\end{array}} \right.,
\end{split}
\label{Theta2_cal}
\end{equation}
and
\begin{equation}
{\Theta _3} = \left\{ {\begin{array}{*{20}{c}}
{0,}&{{\gamma_{th}} \ge \beta }\\
{\underbrace {\begin{array}{*{20}{c}}
{\Pr \left\{ {{Y_i} \ge \frac{{{A_{{\gamma_{th}}}}}}{{{\gamma _0}}},{Z_i} \ge \frac{{{A_{{\gamma_{th}}}}}}{{{\gamma _E}}},} \right.}\\
{\left. \hspace{1cm}{{Y_i} < \frac{{{2^{{R_{s,N}}}}{\gamma _E}{Z_i} + {\gamma _{s,N}}}}{{{\gamma _0}}}} \right\}}
\end{array}}_{{\Lambda _3}},}&{{\gamma_{th}} < \beta }
\end{array}} \right..
\label{Theta3_cal}
\end{equation}

To calculate $\Lambda_2$, we observe that if $\frac{{{A_{{\gamma_{th}}}}}}{{{\gamma _0}}} > \frac{\gamma _{s,N}}{\gamma _0}$ or ${R_{s,N}} < \eta$  then $\Lambda_2 = 0$, otherwise, ${\Lambda _2} = \Pr \left\{ {\frac{{{A_{{\gamma_{th}}}}}}{{{\gamma _0}}} \le {Y_i} < \frac{\gamma _{s,N}}{\gamma _0},{Z_i} < \frac{{{A_{{\gamma_{th}}}}}}{{{\gamma _E}}}} \right\}$. By using the definition of the CDF, $\Lambda_2$ is expressed as
\begin{equation}
{\Lambda _2} = \left\{ {\begin{array}{*{20}{c}}
{0,}&{{R_{s,N}} < \eta }\\
{\begin{array}{*{20}{c}}
{\left[ {{F_{{Y_i}}}\left( {\frac{{{\gamma _{s,N}}}}{{{\gamma _0}}}} \right) - {F_{{Y_i}}}\left( {\frac{{{A_{{\gamma_{th}}}}}}{{{\gamma _0}}}} \right)} \right]}\\
{ \times {F_{{Z_i}}}\left( {\frac{{{A_{{\gamma_{th}}}}}}{{{\gamma _E}}}} \right)\hspace{2cm}}
\end{array},}&{{R_{s,N}} > \eta }
\end{array}} \right.,
\label{Lambda_2}
\end{equation}

Regarding $\Lambda_3$, we observe that if $\frac{{{2^{{R_{s,N}}}}{\gamma _E}{Z_i} + {\gamma _{s,N}}}}{{{\gamma _0}}} < \frac{{{A_{{\gamma_{th}}}}}}{{{\gamma _0}}}$ or ${Z_i} < \frac{{{A_{{\gamma_{th}}}} - {\gamma _{s,N}}}}{{{2^{{R_{s,N}}}}{\gamma _E}}}$  then $\Lambda_3 = 0$, otherwise, ${\Lambda _3} = \Pr \left\{ {\frac{{{A_{{\gamma_{th}}}}}}{{{\gamma _0}}} \le {Y_i} < \frac{{{2^{{R_{s,N}}}}{\gamma _E}{Z_i} + {\gamma _{s,N}}}}{{{\gamma _0}}},{Z_i} \ge \frac{{{A_{{\gamma_{th}}}}}}{{{\gamma _E}}}} \right\}$. Since $\frac{{{A_{{\gamma_{th}}}} - {\gamma _{s,N}}}}{{{2^{{R_{s,N}}}}{\gamma _E}}} < \frac{{{A_{{\gamma_{th}}}}}}{{{\gamma _E}}}$, $\forall {\gamma_{th}} < \beta$, hence $\Lambda_3$ is given by
\begin{equation}
\begin{split}
%\Lambda_3 &= \int\limits_{\frac{{{A_{{\gamma_{th}}}}}}{{{\gamma _E}}}}^\infty  {\int\limits_{\frac{{{A_{{\gamma_{th}}}}}}{{{\gamma _0}}}}^{\frac{{{2^{{R_{s,N}}}}{\gamma _E}{Z_i} + {\gamma _{s,N}}}}{{{\gamma _0}}}} {{f_{{Y_i}}}\left( y \right){f_{{Z_i}}}\left( x \right)dydx} } \\
\Lambda_3 &= \int\limits_{\frac{{{A_{{\gamma_{th}}}}}}{{{\gamma _E}}}}^\infty  {\left[ {{F_{{Y_i}}}\left( {\frac{{{2^{{R_{s,N}}}}{\gamma _E}x + {\gamma _{s,N}}}}{{{\gamma _0}}}} \right) - {F_{{Y_i}}}\left( {\frac{{{A_{{\gamma_{th}}}}}}{{{\gamma _0}}}} \right)} \right]} \\
&\hspace{5.5cm} \times {f_{{Z_i}}}\left( x \right)dx.
\end{split}
\label{Lambda_3}
\end{equation}
From (\ref{Theta1_cal}) to (\ref{Lambda_3}), $SOP_{N,i}$ is obtained as in (\ref{SOPNi_def}) and the proof is completed.

\section{Proof of Theorem 1}
By using {Proposition} 1, $F_{{\gamma_{FS_{\hat i}}^{x_F}},I}\left( g_{x,F} \right)$ is expressed as
\begin{equation}
\begin{split}
&{F_{\gamma _{F{S_{\hat i}}}^{{x_F}},I}}\left( {{g_{x,F}}} \right) \\
&= \left\{ {\begin{array}{*{20}{c}}
{1,}&{x \ge {u_F}}\\
{1 - \sum\limits_{k = 0}^{{a_F -1}} {\frac{1}{{k!\lambda _{FS}^k}}{{\left( {\frac{{{m_F}{A_{{g_{x,F}}}}}}{{{\gamma _0}}}} \right)}^k}{e^{ - \frac{{{m_F}{A_{{g_{x,F}}}}}}{{{\gamma _0}{\lambda _{FS}}}}}}} ,}&{x < {u_F}}
\end{array}} \right..
\end{split}
\label{cdf_gfsgxfI}
\end{equation}

Based on {Proposition} 3 and (\ref{cdf_gfsgxfI}), SOP of user $F$ in (\ref{SOPFi_def}) can be rewritten as
\begin{small}\begin{equation}
\begin{split}
SOP_{F,I} &= \sum\limits_{k = 0}^{{a_E -1}} {\frac{{m_E^k{\alpha _F}}}{{k!\lambda _{ES}^k\gamma _E^k}}\int\limits_0^\beta  {\frac{{A_x^{k - 1}{e^{ - \frac{{{m_E}{A_x}}}{{{\gamma _E}{\lambda _{ES}}}}}}}}{{{{\left( {{\alpha _F} - {\alpha _N}x} \right)}^2}}}\left( {\frac{{{m_E}{A_x}}}{{{\lambda _{ES}}{\gamma _E}}} - k} \right)dx} } \\
& \quad - \sum\limits_{m = 0}^{{a_F - 1}} {\sum\limits_{n = 0}^{{a_E - 1}} {\frac{{{\alpha _F}m_F^mm_E^n}}{{m!n!\lambda _{FS}^m\lambda _{ES}^n\gamma _0^m\gamma _E^n}}} \int\limits_0^{{u_F}} {{h_{F,I}}\left( x \right)dx} } \\
& = 1 - \sum\limits_{m = 0}^{{a_F - 1}} {\sum\limits_{n = 0}^{{a_E - 1}} {\frac{{{\alpha _F}m_F^mm_E^n}}{{m!n!\lambda _{FS}^m\lambda _{ES}^n\gamma _0^m\gamma _E^n}}} \int\limits_0^{{u_F}} {{h_{F,I}}\left( x \right)dx} },
\end{split}
\label{SOPFI_cal}
\end{equation}
\end{small}where, (\ref{SOPFI_cal}) is obtained by the fact that $u_F \le \beta$, $\forall R_{s,F} \ge 0$ and ${h_{F,I}}\left( x \right) = \frac{{A_{{g_{x,F}}}^mA_x^{n - 1}}}{{{{\left( {{\alpha _F} - {\alpha _N}x} \right)}^2}}}{e^{ - \frac{{{m_F}{A_{{g_{x,F}}}}}}{{{\gamma _0}{\lambda _{FS}}}} - \frac{{{m_E}{A_x}}}{{{\gamma _E}{\lambda _{ES}}}}}}\left( {\frac{{{m_E}{A_x}}}{{{\gamma _E}{\lambda _{ES}}}} - n} \right)$. Note that it is challenging to resolve the integral in (\ref{SOPFI_cal}), hence an approximation solution is proposed by using the Gaussian-Chebyshev quadrature \cite{Hil1987} as follows:
\begin{equation}
\int\limits_0^{{u_F}} {{h_{F,I}}\left( x \right)dx}  \approx \frac{{\pi {u_F}}}{{2N}}\sum\limits_{i = 0}^N {h_{F,I} \left[ {\frac{{\left( {{v_i} + 1} \right){u_F}}}{2}} \right]\sqrt {1 - v_i^2} }.
\label{int_hFI_appro}
\end{equation}
By substituting (\ref{int_hFI_appro}) into (\ref{SOPFI_cal}), $SOP_{F,I}$ is obtained as in (\ref{SOPFI_final}). The proof of Theorem 1 is completed.

\section{Proof of Theorem 2}
From (\ref{SOPNi_def}), $SOP$ of user $N$ in Solution I is expressed as in (\ref{SOPNI_final}), in which $\Lambda_{1,I}$ and $\Lambda_{2,I}$ is obtained by substituting $F_{{Y_{\hat i}},I}\left( x \right)$ in (\ref{cdf_hnsI_final}) and $F_{Z_{\hat i}}\left( x \right)$ in (\ref{cdf_hes_final}) into $\Lambda_{1}$ and $\Lambda_{2}$ in (\ref{SOPNi_def}). To derive $\Lambda_{3,I}$, we rewrite $\Lambda_{3,I}$ by substituting $F_{{Y_{\hat i}},I}\left( x \right)$ in (\ref{cdf_hnsI_final}) and $f_{Z_{\hat i}}\left( x \right)$ in (\ref{pdf_hes_final}) into $\Lambda_3$ in (\ref{SOPNi_def}) as follows:
\begin{small}\begin{equation}
\begin{split}
\Lambda_{3,I} &= \underbrace {\left[ {1 - {F_{{Y_{\hat i}},I}}\left( {\frac{{{A_{{\gamma_{th}}}}}}{{{\gamma _0}}}} \right)} \right]\left[ {1 - {F_{{Z_{\hat i}}}}\left( {\frac{{{A_{{\gamma_{th}}}}}}{{{\gamma _E}}}} \right)} \right]}_{B_{N,I}^{(1)}} \\
& \quad + \sum\limits_{p = 1}^{{L_S}} {\sum\limits_{{\Delta _N} = p} {\frac{{{\Phi _N}m_E^{{a_E}}}}{{\Gamma \left( {{m_E}{L_E}} \right)\lambda _{ES}^{{a_E}}}}} } \int\limits_{\frac{{{A_{{\gamma_{th}}}}}}{{{\gamma _E}}}}^\infty  {{{\left( {\frac{{{2^{{R_{s,N}}}}{\gamma _E}x + {\gamma _{s,N}}}}{{{\gamma _0}}}} \right)}^{{\varphi _N}}}} \\
& \quad \times {x^{{a_E} - 1}}{e^{ - \frac{{p{m_N}\left( {{2^{{R_{s,N}}}}{\gamma _E}x + {\gamma _{s,N}}} \right)}}{{{\gamma _0}{\lambda _{NS}}}} - \frac{{{m_E}x}}{{{\lambda _{ES}}}}}}dx.
\end{split}
\label{Lambda3I_cal}
\end{equation}\end{small}By applying binomial expansion for ${\left( {{2^{{R_{s,N}}}}{\gamma _E}x + {\gamma _{s,N}}} \right)^{{\varphi _N}}}$ and upper incomplete Gamma function \cite[Eq. 8.350.2]{Grad2007} into (\ref{Lambda3I_cal}), $\Lambda_{3,I}$ is obtained as in (\ref{SOPNI_final}). The proof of Theorem II is completed.

%\section*{Acknowledgment}
%\textcolor{red}{This research is funded by Vietnam National Foundation for Science and Technology Development (NAFOSTED) under grant number 102.04-2017.301.}


\ifCLASSOPTIONcaptionsoff
  \newpage
\fi



%\bibliographystyle{IEEEtran}
\bibliographystyle{IEEEtran}
\bibliography{IEEEabrv,dung}

%\begin{biography}[{\includegraphics[width=1in,height=1.25in,clip,keepaspectratio]{mshell}}]{Michael Shell}

%\end{IEEEbiography}
%% insert where needed to balance the two columns on the last page with
%%\newpage


%\vfill


\end{document}